%Paper: hep-th/9404041
%From: Hidetoshi Awata <awata@ps1.yukawa.kyoto-u.ac.jp>
%Date: Fri, 08 Apr 1994 01:20:45 +0900

%%%%%%%%%%%%%%%%%%%%%%%%%%%%% Figures %%%%%%%%%%%%%%%%%%%%%%%%%%%%%%%%%%
%
%  This file has 2 postscript files after the line ***********************.
%  Please save them named swfig1.eps and swfig2.eps.
\input epsf.tex
%%%%%%%%%%%%%%%%%%%%%%%%%%%%%%%%%%%%%%%%%%%%%%%%%%%%%%%%%%%%%%%%%%%%%%%%%%
%
%
%   Quasifinite Highest Weight Modules over Super $W_{1+\infty}$ Algebra
%
%
%           H. Awata, M. Fukuma, Y. Matsuo and S. Odake
%
%                       Plain Tex file
%
%%%%%%%%%%%%%%%%%%%%%%%%%%%% font %%%%%%%%%%%%%%%%%%%%%%%%%%%%%%%%%%%%%
\font\bigrm=cmb10 scaled\magstep1
\font\srm=cmr9
\font\csc=cmcsc10
\font\tennn=msbm10
\font\twelvenn=msbm10 scaled\magstep1
\def\Sbm#1{\leavevmode\raise-.25ex\hbox{\twelvenn #1}}
\def\sbm#1{\leavevmode\raise-.15ex\hbox{\tennn #1}}
%%%%%%%%%%%%%%%%%%%%%%% footnote %%%%%%%%%%%%%%%%%%%%%%%%%%%%%%%%%%%

\def\foot{\baselineskip=12pt\srm\footnote}

%%%%%%%%%%%%%%%%%%%%%%% style %%%%%%%%%%%%%%%%%%%%%%%%%%%%%%%%%%%%%%%%%%%%

\magnification=\magstep1
%\hoffset-.27truein
\vsize 8.9truein    \hsize 6.3truein %\hsize 6.8truein
\topskip 10pt       \leftskip 0pt  \rightskip 0pt
\baselineskip=19pt plus 2pt minus 1pt
\parskip 0pt plus 1pt    \parindent 20pt

%%%%%%%%%%%%%%%%%%%%%%%%%% def %%%%%%%%%%%%%%%%%%%%%%%%%%%%%%%%%%%%%%%%%

\def\ha{{1 \over 2}}
\def\->{\rightarrow}       \def\<-{\leftarrow}
\def\<{\langle\,}          \def\>{\,\rangle}
\def\[{\left [\,}          \def\]{\,\right ]}
\def\({\left (\,}          \def\){\,\right )}
\def\mBRA{\left\{\, }      \def\mKET{\,\right\} }

\def\|{\vert}

\def\Smat#1{\left[\matrix{#1}\right]}

\def\der#1{{\partial \over \partial #1}}

\def\no{\noindent}

\def\qed{\ \ \ \ \hbox{\hfill\vbox{\hrule width8pt
    \noindent\vrule height8pt\hskip 7.5pt\vrule height8pt\hrule width8pt}}}
\def\set#1#2{\left\{\,\left.#1\,\right\vert\,#2\,\right\}}
\def\tsum_#1^#2{{\textstyle\sum_{#1}^{#2}}}
\def\tsum^#1_#2{{\textstyle\sum^{#1}_{#2}}}
\def\tsum_#1{{\textstyle\sum_{#1}}}

%%%%%%%%%%%%%%%%%%% Greek Letters %%%%%%%%%%%%%%%%%%%%%%%%%%%%%%%%%%%%%

\def\a{\alpha}
\def\b{\beta}
\def\d{\delta}	    \def\D{\Delta}
\def\e{\epsilon}
\def\g{\gamma}	    

\def\la{\lambda}    \def\La{\Lambda}
      
\def\p{\phi}              \def\vp{\varphi}

\def\sig{\sigma}    
\def\t{\theta}	        

\def\l{\ell}

%%%%%%%%%%%%%%%%%%%%%%%% Bold %%%%%%%%%%%%%%%%%%%%%%%%%%%%%%%%%%%%%%%%%

\def\bZ{{\bf  Z}} \def\bC{{\bf  C}} 
\def\cA{{\cal A}}   
  \def\cS{{\cal S}} \def\cW{{\cal W}}
\def\tA{\hat A}   \def\tC{\hat C}
\def\tW{\widetilde W} \def\tPsi{\widetilde\Psi}
\def\cg{{\cal G}}
\def\bZ{{\sbm Z}} \def\bC{{\sbm C}}
%%%%%%%%%%%%%%% definition for this article %%%%%%%%%%%%%%%%%%%%%%%%%%%%

\def\tAs{\tilde\cA_{\sig         } }
\def\hAs{ \hat \cA_{\sig,\,\str  } }
\def\hAsz{\hat \cA_{\sig,\,\str_0} }

\def\PC{ \Psi_{\sig,\,\str  }}
\def\PCz{\Psi_{\sig,\,\str_0}}

\def\Win{\cW_{\infty}}
\def\Winf{\cW_{1+\infty}}
\def\swinf{{\it sw}_{1+\infty}}
\def\SWinf{\cS\cW_{1+\infty}}
\def\h->{\hookrightarrow }
\def\gl{{\rm gl(\infty)}}
\def\glinf{{\rm gl(\infty\vert\infty)}}
\def\Glinf{{\rm \widehat{gl}(\infty\vert\infty)}}

\def\AA{A=0,1,\pm}
\def\z{[z,z^{-1}]}

\def\str{{\rm str}}

\def\der#1{\partial_#1}

\def\Emat{\Smat{\e^0 &\e^+ \cr
                \e^- &\e^1 \cr}}
\def\Fmat#1#2{\Smat{#1^0(#2) & #1^+(#2)\cr
                    #1^-(#2) & #1^1(#2)\cr}}
\def\Fgmat#1#2{\Smat{#1^1(#2) & #1^-(#2)\cr
                     #1^+(#2) & #1^0(#2)\cr}}
\def\aBmat#1{\Smat{\a b^0_{#1}(D) &\b b^+_{#1}(D)\cr
                   \g b^-_{#1}(D) &\d b^1_{#1}(D)\cr}}
\def\BR{\break}

%%%%%%%%%%%%%%%%%%% Adress %%%%%%%%%%%%%%%%%%%%%%%%%%%%%%%%%%%%%%%%%%%%
\def\Fha{e-mail address : awata@yukawa.kyoto-u.ac.jp;
Address after April 1, 1994:
Research Institute for Mathematical Sciences,
Kyoto University, Kyoto 606, Japan}
\def\Fmf{e-mail address : fukuma@yukawa.kyoto-u.ac.jp}
\def\Fym{e-mail address : matsuo@danjuro.phys.s.u-tokyo.ac.jp;
Address after April 1, 1994:
Uji Research Center, Yukawa Institute for Theoretical Physics,
Kyoto University, Uji 611, Japan}
\def\Fso{e-mail address : odake@jpnyitp.yukawa.kyoto-u.ac.jp}
%%%%%%%%%%%%%%%%%%%%%%% Tytle %%%%%%%%%%%%%%%%%%%%%%%
{\baselineskip=12pt
\rightline{\vbox{\hbox{YITP/K-1055}
                 \hbox{UT-670}
                 \hbox{SULDP-1994-2}
                 \hbox{March 1994}
}}}

\vskip.35in
\centerline
{\bigrm  Quasifinite Highest Weight Modules over the Super $\Winf$ Algebra}
\vskip.45in\centerline{
{\csc Hidetoshi AWATA}{\foot{$^1$}\Fha} ,
{\csc Masafumi FUKUMA}{\foot{$^2$}\Fmf} ,
{\csc Yutaka   MATSUO}{\foot{$^3$}\Fym}}
\centerline{\ and \
{\csc Satoru    ODAKE}{\foot{$^4$}\Fso}
}\def\csc{}

{\baselineskip=14pt
\it\vskip.25in
\centerline{$^{1,2}$ Yukawa Institute for Theoretical Physics}
\centerline{Kyoto University, Kyoto 606, Japan}
\vskip.1in
\centerline{$^3$ Department of Physics, University of Tokyo}
\centerline{Bunkyo-ku, Hongo 7-3-1, Tokyo 113, Japan}
\vskip.1in
\centerline{$^4$ Department of Physics, Faculty of Liberal Arts}
\centerline{Shinshu University, Matsumoto 390, Japan}
}
\vskip.35in\centerline{\bf Abstract}\vskip.15in
We study quasifinite highest weight modules over
the supersymmetric extension of the $\Winf$ algebra
on the basis of the analysis by Kac and Radul.
We find that the quasifiniteness of the modules is again
characterized by polynomials,
and obtain the differential equations for highest weights.
The spectral flow, free field realization over the $(B,C)$--system,
and the embedding into $\Glinf$ are also presented.

\vskip.1in
hep-th/9404041
\vfill\eject

%%%%%%%%%%%%%%%%%%%%%%%%%%%%%%%%%%%%%%%%%%%%%%
%%                                          %%
%%  1. Introduction                         %%
%%                                          %%
%%%%%%%%%%%%%%%%%%%%%%%%%%%%%%%%%%%%%%%%%%%%%%%%%%%%%%%%%%%%%%%%%%%%%%%
\vskip 3mm
\no{\bf  1. Introduction }
\vskip2mm
%%%%%%%%%%%%%%%%%%%%%%%% Introduction %%%%%%%%%%%%%%%%%%%%%%%%%%%%%%%
\no
Conformal field theory has attracted much interest for the last ten years,
since it describes classical vacua of string theory and
two--dimensional statistical system
at fixed points of the renormalization group.
The representation theory of the Virasoro algebra plays a central role [BPZ].
However, when the systems have larger symmetries,
the Virasoro algebra must be extended.
For example, when supersymmetry exists one will be led to
the super Virasoro algebra [NS,\ R,\ GS],
while for ${\bZ}_N$ symmetry what is called the $\cW_N$ algebra
will be relevant [Z,\ BS].

In the $\cW_N$ algebra (or its supersymmetric extension [IMY]),
there are $(N-1)$ generating currents with spins $s=2,3,...,N$
(and their superpartners if supersymmetry exists).
Here $s=2$ corresponds to the energy momentum tensor.
The peculiar nature of the algebra is in its nonlinearity, {\it i.e.},
the singular part of the operator product of generating currents
is not expanded as a linear combination of the generating currents,
and one has to introduce composite fields made of the currents.
The occurrence of such operators implies that
the corresponding algebra is not a Lie algebra in ordinary sense.
Indeed, the check of the Jacobi identity gives severe
restrictions on the structure constants.

The situation changes drastically
if we take a suitable limit $N\->\infty$ [B].
The resulting algebra, called the $\Win$ algebra,
becomes a Lie algebra [PRS1].
This limiting procedure is essentially equal to regarding
the composite fields
needed in the operator product expansion of lower spin currents,
as a new generating current with higher spin.

Great simplification further occurs if we add spin--1 current
($u(1)$ current) to the $\Win$ algebra [PRS2].
The obtained algebra is named the $\Winf$ algebra for this historical reason.
Although there are several types of $\cW$ infinity--like algebras [BK,\ FFZ],
we may say, at cost of rigor, that this is the most fundamental one.
All other algebras such as $\Win$ and $\cW_N$ are obtained
by imposing some suitable constraints on it.

The $\Winf$ algebra naturally arises in various physical systems.
Firstly, in two--dimensional quantum gravity
(the square root of) the generating function of scaling operators
is identified with a $\tau$--function of KP hierarchy and
obeys the vacuum condition of the $\Winf$ algebra
[FKN,\ DVV,\ IM,\ KS,\ S,\ G].
Secondly, in the quantum Hall effect,
edge states satisfy the highest weight condition of the $\Winf$ algebra,
reflecting the incompressibility of quantum fluid [IKS,\ CTZ].
Some interesting applications are also known in higher dimensional physics,
such as the construction of gravitational instantons [T,\ YC,\ P].
Furthermore, the $\Winf$ algebra is known to be closely related to the
central extension of $\gl$ algebra [KR].
The application to the large $N$ two--dimensional QCD [GT,\ DLS]
seems also intriguing in this context.

The major reason of such generality of the $\Winf$ algebra is that
it is a central extension of the Lie algebra of differential operators
on the circle [PRS1].
Recently, Kac and Radul gave a general framework on such Lie algebra
and classified all the quasifinite representations [KR].
Since the purpose of this paper is to extend
their work to the system with supersymmetry,
it may be instructive to review their main results.

%%%%%%%%%%%%%%%% Kac-Radul %%%%%%%%%%%%%%%%%%%%%%%%%%%%%%%%%%%%%%%

Let $\cg$ be the Lie algebra of differential operators on
the circle; $\cg\! =\!\set{\! z^n f(D)}{ n\!\in\bZ\!}$, where
$f(w)\in\bC[w]$ (polynomial ring with $w$ indeterminate) and
$D\equiv z{d \over dz}$.
Let then $\Winf$ be the central extension of $\cg$,
and for $z^nf(D) \in \cg$ we denote the corresponding
operator in $\Winf$ by $W(z^nf(D))$.
The central extension is defined by the following commutation relations:
$$
{}~\left[ W(z^nf(D)),\,W(z^mg(D))\right]
=W\(\left[ z^nf(D),\,z^mg(D)\right]\)\,+\,C\,\Psi(z^n f(D),z^m g(D)),
$$
where the two--cocycle $\Psi$ is given by
$$\eqalign{
 \Psi(z^nf(D), z^mg(D)) &= - \Psi(z^mg(D),z^nf(D))\cr
&=\left\{\matrix{
 \sum_{j=1}^n f(-j)g(n-j)~~~&{\rm if}~n=-m>0 \hfill\cr
 0                       ~~~&{\rm if}~n+m\neq 0~\quad{\rm or}\quad~n=m=0.\cr
}\right.
}$$
More symmetrically, it is written as
$$
{}~\left[W(z^ne^{x D}),\,W(z^me^{y D})\right]
\,=\,\(e^{x m}-e^{y n}\)\,W(z^{n+m}e^{(x+y)D})
\,-\,C\,\d_{n+m,0}\,
{{e^{xm}-e^{yn}}\over{e^{x+y}-1}}\,.
$$
The two--cocycle is shown to be unique up to coboundaries [Li,\ F].

The $\Winf$ algebra has the following principal gradation:
$$\eqalign{
\Winf&=\bigoplus_{n\in{\bZ}}\(\Winf\)_n,\cr
\(\Winf\)_n&=\set{z^n f(D)}{f(w)\in\bC[w]}.
}$$
Note that the Cartan subalgebra is given by
$\(\Winf\)_0=\bigoplus_{s=0}^\infty \bC W(D^s)$.
A highest weight state $|\la\>$ is thus characterized by the condition
$$\eqalign{
\(\Winf\)_n|\la\>&=0~~~(n\geq1), \cr
\(\Winf\)_0|\la\>&\subset\bC|\la\>.
}$$
We introduce the energy operator $L_0\equiv-W(D)$ and call its
eigenvalue of state the energy level.
Note that $[L_0, W(z^{-k}f(D))]\,=\,+\,k\,W(z^{-k}f(D))$.

%%%%%%%%%%%%%%%% quasifinite representation %%%%%%%%%%%%%%%%%%%%

At each energy level $k$, there might be infinitely many states,
reflecting the infinitely many degrees of freedom in the polynomial ring.
The {\it quasifinite} representation is obtained if we require
that all but finitely many states at each energy level vanish.
More precisely, it is equivalent to saying that the set
$$
I_{-k}\equiv
\set{f(w)\in\bC[w]}{W(z^{-k}f(D))|\la\>=0}
$$
is different from $\{0\}$ for any $k\geq1$.
Since $I_{-k}$ is an ideal in $\bC[w]$,
we can introduce the monic (with unit leading coefficient)
generating polynomial $b_k(w)$; $I_{-k}=(b_k(w))$.
These polynomials $\{b_k(w)\}_{k=1,2,3,...}$
are called characteristic polynomials.

A surprising result obtained in ref. [KR] is that they are
almost uniquely determined by the first characteristic polynomial
$b(w) \equiv b_1(w)$.
To show this, one has to observe that
(i) $b_k(w)$ is divided by $\l.c.m.(b(w),b(w-1),...,b(w-k+1))$,
(ii) $b(w)b(w-1)...b(w-k+1)$ is divided by $b_k(w)$.
These statements are proved by using the null state
conditions of the $\Winf$ algebra.
Thus, if difference of any two distinct roots of $b(w)$ is not an integer,
then $b_k(w)$ is uniquely expressed as $b_k(w)=b(w)b(w-1)...b(w-k+1)$.

It is further shown that the generating function $\D(x)$ of highest weights:
$$
\D(x)\equiv-\sum_{s=0}^\infty {x^s \over s!}\D^{(s)}
\quad{\rm for}\quad W(D^s)|\la\> = \D^{(s)}|\la\>
$$
satisfies a simple differential equation:
$$
b\({d\over dx}\)\[(e^x-1)\D(x)+C\]\,=\,0.
$$

%%%%%%%%%%%%%%% super W %%%%%%%%%%%%%%%%%%%%%%%%%%%%

To cover all $\cW$--like algebras with supersymmetry, the $\Winf$
algebra must be extended such as to contain supersymmetry.
Such extension was first considered in ref.\ [MR,\ UY]
in the context of supersymmetric Kadmtsev-Petviashvili hierarchy,
and also in ref.\ [BdWV,\ BPRSS],
where explicit form of (anti--) commutation relations are given.
In this paper, we reformulate their work on the super $\Winf$ algebra
(later denoted by $\SWinf$) and develop the representation theory,
on the basis of the analysis by Kac and Radul for the $\Winf$ algebra.
We find that quasifiniteness is again characterized by polynomials,
and that the highest weights are expressed
in terms of combined differential equations.

%%%%%%%%%%%%%%%%%%%%%%%%%%%%%%%%%%%%%%%%%%%%%%%%%%%%

The present paper is organized as follows.
In sect.\ 2 and sect.\ 3,
we discuss the general theory of the super $\Winf$ algebra,
viewing it as a central extension of the Lie superalgebra
of superdifferential operators on the circle.
In sect.\ 4, we classify the quasifinite highest weight representations
of the super $\Winf$ algebra, and then, in sect.\ 5,
derive the differential equation which determines the highest weights.
In sect.\ 6, we discuss the spectral flow
(two--parameter family of automorphisms) in $\SWinf$.
In sect.\ 7, we consider the $(B,C)$--system as an example.
Sect.\ 8 is devoted to conclusion and discussion.
The embedding of $\SWinf$ into $\Glinf$
and null vector condition are given in Appendices.

%%%%%%%%%%%%%%%%%%%%%%%%%%%%%%%%%%%%%%%%%%%%%%%%%%%
%%                                               %%
%%  2. General Theory of Lie Superalgebra of     %%
%%     Superdifferential Operators on the Circle %%
%%                                               %%
%%%%%%%%%%%%%%%%%%%%%%%%%%%%%%%%%%%%%%%%%%%%%%%%%%%%%%%%%%%%%%%%%%%%%%%%%%
\bigbreak
\no{\bf 2. General Theory of Lie Superalgebra of
           Superdifferential Operators on the Circle}
\vskip2mm
%%%%%%%%%%%%%%%%%%%%%%%%%%%%%%%%%
%%                             %%
%%  2.1                        %%
%%                             %%
%%%%%%%%%%%%%%%%%%%%%%%%%%%%%%%%%
%%%%%%%%%%%%%% algebra $\A$ and Laurent polynomial algebra %%%%%%%%%%%%%%%%
\no{\bf 2.1.}~
Let $\cA=\cA^{(0)}\oplus\cA^{(1)}$ be a $\bZ_2$--graded associative algebra
and let $\sig\,:\,\cA\->\cA$ be a $\bZ_2$--preserving automorphism of $\cA$;
$\sig\(\cA^{(0,1)}\)=\cA^{(0,1)}$.
We specify the $\bZ_2$--gradation of an element
$a\in\cA^{(0)}$ ({\it resp.\ }$a\in\cA^{(1)}$) as
$|a|=0$ ({\it resp.\ }$|a|=1$).
We then introduce the {\it twisted Laurent polynomial algebra}
$\cA\z$ over $\cA$:
$$\eqalign{
\cA\z&\equiv \bC\z\,\otimes_\bC\,\cA \cr
&=\set{\sum_{n\in\bZ}z^n\otimes a_n}
{a_n\in\cA,\,{\rm~all~but~finite~number~of~} a_n \hbox{{\rm 's~vanish}}}
}$$
with the following $*$--multiplication:
$$
(z^n\otimes a)*(z^m\otimes b)\equiv z^{n+m}\otimes \sig^m(a)\cdot b.
$$
Note that
the $\bZ_2$--gradation of $\cA$  naturally induces that of $\cA\z$;
$|z^n\otimes a|\equiv|a|$ if $a\in\cA^{(0)}$ or $\cA^{(1)}$.
In what follows, we will denote $z^n\otimes a$ by $z^n a$ for simplicity.

%%%%%%%%%%%%%%%%%%%%%%%%%%%%%%%%%
%%                             %%
%%  2.2                        %%
%%                             %%
%%%%%%%%%%%%%%%%%%%%%%%%%%%%%%%%%
%%%%%%%%%%%%%%%%%% Lie algebra $\tAs$ %%%%%%%%%%%%%%%%%%%%%%%%%%%%%%%%%
\no{\bf 2.2.}~
Let $\tAs$ denote the algebra $\cA\z$ regarded as
a Lie superalgebra with respect to the usual (anti--) bracket:
$$\eqalign{
\[z^n a,z^m b\mKET &\equiv (z^n a)*(z^m b)-(-1)^{|a||b|}(z^m b)*(z^n a)\cr
&=z^{n+m}\sig^m(a)\cdot b -(-1)^{|a||b|}z^{n+m}\sig^n(b)\cdot a.
}$$

%%%%%%%%%%%%%%%%%%%%%%%%%%%%%%%%%
%%                             %%
%%  2.3                        %%
%%                             %%
%%%%%%%%%%%%%%%%%%%%%%%%%%%%%%%%%
%%%%%%%%%%%%%%%%% affine Lie algebra $hAs$ %%%%%%%%%%%%%%%%%%%%%%
\no{\bf 2.3.}~
Fix a linear map $\str \,:\,\cA\->V$ such that
$\str\,ab=(-1)^{|a||b|}\,\str\,ba$,
where $V$ is a vector space over $\bC$.
Then we can define a {\it central extension}
 $\hAs$ of $\tAs$ by $V$,
$0 \-> V \-> \hAs \-> \tAs \-> 0$, as follows.
First, we notice that the map
$\PC\,:\,\tAs\times\tAs\->V$ defined by
$$\eqalign{
&\PC(z^n a,z^m b) \cr
&~~~~\equiv -(-1)^{|a||b|}\PC(z^m b,z^n a)\cr
&~~~~\equiv \left\{\matrix{
\str\((1+\sig+\cdots+\sig^{n-1})(\sig^{-n}(a)\cdot b)\)
&{\rm if }\quad n=-m>0,\cr
0\hfill
&{\rm if }\quad n+m\neq 0 ~{\rm or }~n=m=0,\cr
}\right.
}$$
satisfies the {\it $2$--supercocycle condition}:
$$\eqalign{
{\rm (1)}&~~ \PC(A,B)=-(-1)^{|A||B|} \PC(B,A),\cr
{\rm (2)}&~~ (-1)^{|A||C|} \PC\([A,B\},C\)+
{\rm cyclic~permutation}=0.
\qquad\qquad\qquad\qquad{ }}$$
Thus, denoting by $W(A)$ the element in $\hAs$
which corresponds to $A\in\tAs$,
we define the (anti--) bracket of two elements
$W(A),\,W(B)\,\in\,\hAs$ by the following formula:
$$
\[W(A),W(B)\mKET\equiv W(\[A,B\mKET)+ \PC(A,B).
$$
Hereafter, we will restrict ourselves to
one--dimensional central extensions; $V=\bC$.

%%%%%%%%%%%%%%%%%%%%%%%%%%%%%%%%%%%%%%%%%%%%%%
%%                                          %%
%%  3. The Super $\Winf$ Algebra $\SWinf$   %%
%%                                          %%
%%%%%%%%%%%%%%%%%%%%%%%%%%%%%%%%%%%%%%%%%%%%%%%%%%%%%%%%%%%%%%%%%%%
\bigbreak
\no{\bf 3. The Super $\Winf$ Algebra $\SWinf$}
\vskip2mm
%%%%%%%%%%%%%%%%%%%%%% algebra $\cA$ %%%%%%%%%%%%%%%%%%%%%%%%%%%%%%
\def\EA{3.1}
\def\Ecri{3.2}
\def\Ecrii{3.3}
%%%%%%%%%%%%%%%%%%%%%%%%%%%%%%%%%
%%                             %%
%%  3.1                        %%
%%                             %%
%%%%%%%%%%%%%%%%%%%%%%%%%%%%%%%%%
\no{\bf 3.1.}~
In the rest of the present paper, we will exclusively consider the
case where $\cA$ is the polynomial
algebra over $(2\times 2)$ supermatrices:
$$\eqalign{
\cA&\equiv\set{\Fmat{f}{w}}{f^A(w)\in\bC[w]\,;\,\AA }\cr
&=\cA^{(0)}\oplus\cA^{(1)}.
}\eqno(\EA)$$
Here we assign the $\bZ_2$--gradation as follows:
$$\eqalign{
\cA^{(0)}&=\mBRA \Smat{f^0(w) &     0 \cr     0  & f^1(w)\cr}\mKET\cr
\cA^{(1)}&=\mBRA \Smat{    0  & f^+(w)\cr f^-(w) &     0 \cr}\mKET
}\quad\eqalign{
&:\quad \bZ_2{\rm -even},\cr
&:\quad \bZ_2{\rm -odd}.\cr
}$$
Introducing a basis $P_A$ $(\AA)$ in $\cA$ as
$$\eqalign{
P_0 &= \Smat{ 1 & 0 \cr 0 & 0 \cr},\cr
P_+ &= \Smat{ 0 & 1 \cr 0 & 0 \cr},
}\quad\eqalign{
P_1 &= \Smat{ 0 & 0 \cr 0 & 1 \cr}\cr
P_- &= \Smat{ 0 & 0 \cr 1 & 0 \cr}
}\,\eqalign{
&\in\,\cA^{(0)},\cr
&\in\,\cA^{(1)},
}$$
we may denote $F\in\cA$ as $F(w)=f^A(w)P_A$.
Note that the multiplication as matrices respects the $\bZ_2$--gradation.

%%%%%%%%%%%%%%%%%%%%%%%%%%%%%%%%%
%%                             %%
%%  3.2                        %%
%%                             %%
%%%%%%%%%%%%%%%%%%%%%%%%%%%%%%%%%
%%%%%%%%%%%%%%%% Laurent polynomial algebra %%%%%%%%%%%%%%%%%%%%%%%%%%
\def\Fsig{
This choice of $\sig$ is not unique.
In ref.\ [KR], for example, they also consider the case, $\sig'(F(w))=F(qw)$.}
%%%%%%%%%%%%%%%%%%%%
\no{\bf 3.2.}~
Following the general prescription given in the previous section,
we fix a $\bZ_2$--preserving automorphism $\sig\,:\,\cA\->\cA$,
and define a new $\bZ_2$--graded associative algebra
$$\eqalign{
\cA\z&\equiv \bC\z\,\otimes_\bC\,\cA \cr
&=\set{\sum_{n\in\bZ}z^n F_n(w)}
{F_n(w)\in\cA\,,{\rm~all~but~finite~number~of~}F_n(w)\hbox{{\rm 's~vanish}}}
}$$
with the following $*$--multiplication:
$$
(z^n F(w))*(z^m G(w))\equiv z^{n+m}\sig^m\(F(w)\)\cdot G(w).
$$
%Recall that $\cA\z$ has a natural $\bZ_2$--gradation.

We will set $\sig$ as $\sig(F(w))=\sig(f^A(w)P_A)\equiv f^A(w+1)P_A$,
so that we may replace $f^A(w)$ by $f^A(D)$ with $D=z\partial/\partial z$,
and $*$--multiplication by the usual
multiplication as matrices.{\foot{$^\dagger$}\Fsig}
Note here that $f(D)\,z^m=z^mf(D+m)$ for any holomorphic function $f(w)$.

%%%%%%%%%%%%%%%%%%%%%%%%%%%%%%%%%
%%                             %%
%%  3.3                        %%
%%                             %%
%%%%%%%%%%%%%%%%%%%%%%%%%%%%%%%%%
%%%%%%%%%%%%%%%% Lie algebra $\tAs$ %%%%%%%%%%%%%%%%%%%%%%%%%%%%%%%%%
\no{\bf 3.3.}~
Let $\tAs\equiv\swinf$ denote the algebra $\cA\z$
regarded as a Lie superalgebra with the following (anti--) bracket:
$$\eqalign{
&\[z^n F(D),z^m G(D)\mKET=\[z^n f^A(D)P_A,z^m g^B(D)P_B\mKET\cr
&~~~~~~~~\equiv (z^n f^A(D)P_A)\cdot(z^m g^B(D)P_B)
-(-1)^{|P_A||P_B|}(z^m g^B(D)P_B)\cdot(z^n f^A(D)P_A).\cr
}$$

%%%%%%%%%%%%%%%%%%%%%%%%%%%%%%%%%
%%                             %%
%%  3.4                        %%
%%                             %%
%%%%%%%%%%%%%%%%%%%%%%%%%%%%%%%%%
%%%%%%%%%%%%%%%%% affine Lie algebra $\hAsz$ %%%%%%%%%%%%%%%%%%%%%%%%%%%%%%%
\no{\bf 3.4.}~
We now introduce a linear map $\str_0 \,:\,\cA\->\bC$
as $\str_0 F(D)\equiv \str F(0)$, {\it i.e.},
$$
\str_0\Fmat{f}{D}\equiv f^0(0)-f^1(0).
$$
We should notice that $\str_0$ has the following property:
$$
\str_0 F(D)G(D)=(-1)^{|F(D)||G(D)|}\str_0 G(D)F(D).
$$
Thus, we can define
a one--dimensional central extension $\hAsz\equiv\SWinf$ of $\tAs=\swinf$
through the following (anti--) commutation relation:
$$
\[W(z^n F(D)),W(z^m G(D))\mKET\equiv W(\[z^n F(D),z^m G(D)\mKET)
-C \PCz(z^n F(D),z^m G(D)),
\eqno(\Ecri)$$
where $C$ is the central charge,
and for $n=-m>0$, the $2$--supercocycle $\PCz$ is given by
$$\eqalign{
\PCz(z^n F(D),z^m G(D))
&=\str_0\((1+\sig+\cdots+\sig^{n-1})(\sig^{-n}(F(D))\cdot G(D))\)\cr
&=\sum_{j=1}^n f^A(-j) g^B(n-j) \,\str\,P_A P_B\cr
&=\sum_{j=1}^n \{ f^0(-j) g^0(n-j) + f^+(-j) g^-(n-j)\cr
&\qquad\quad    - f^-(-j) g^+(n-j) - f^1(-j) g^1(n-j)\}.
}$$
Note here that
$$
\str\, P_A P_B = \mBRA
\matrix{
\hfill 1 &~~~{\rm if }~~(A,B)=(0,0)~~{\rm or }~~(+,-)\cr
      -1 &~~~{\rm if }~~(A,B)=(-,+)~~{\rm or }~~(1,1)\cr
\hfill 0 &~~~{\rm otherwise.}\hfill\cr
}\right.
$$
The above (anti--) commutation relations can be rewritten in a simpler form
if we introduce $z^n e^{xD}$ as a generating series for $z^nD^k$:
$$\eqalign{
  \lbrack W (z^n &e^{xD} P_A),W(z^m e^{yD} P_B) \rbrace \cr
&=                    e^{xm}\,W(z^{n+m} e^{(x+y)D} P_A P_B)
  -(-1)^{|P_A||P_B|}\,e^{yn}\,W(z^{n+m} e^{(x+y)D} P_B P_A) \cr
&~\hskip2.5truein
+ C\,{e^{xm}-e^{yn}\over e^{x+y}-1}\,\d_{n+m,0}\,\str P_A P_B.
}\eqno{(\Ecrii)}$$
We remark that the indices $n$ and $m$ need not be integers in
this expression.

The bosonic part of this algebra is the direct sum of
two $\Winf$ algebras with central charges $C$ and $-C$.

%%%%%%%%%%%%%%%%%%%%%%%%%%%%%%%%%
%%                             %%
%%  3.5                        %%
%%                             %%
%%%%%%%%%%%%%%%%%%%%%%%%%%%%%%%%%
%%%%%%%%%%%%%%%%%%% principal gradiation %%%%%%%%%%%%%%%%%%%%%%%%
\no{\bf 3.5.}~
The {\it principal gradation} in $\SWinf$ may be introduced with half--integer
labels $\a\in\bZ/2$ as
$\SWinf=\bigoplus_{\a\in\bZ/2}\(\SWinf\)_\a$,
where
$$\eqalign{
\(\SWinf\)_{\a=n}&\equiv
\set{\ W{\(z^n\(f^0(D)P_0+f^1(D)P_1\)\) }}{f^{0,1}(w)\in\bC[w]},\cr
\(\SWinf\)_{\a=n+1/2}&\equiv
\set{\ W{\(z^n f^+(D)P_+ +z^{n+1}f^-(D)P_-\)}}{f^\pm(w)\in\bC[w]}.
}$$
In fact, one can easily show that
$\[\(\SWinf\)_\a,\(\SWinf\)_\b\mKET\subset\(\SWinf\)_{\a+\b}$
with $\a,\b\in\bZ/2$.
We notice that
the Cartan subalgebra of $\SWinf$ is given by $\(\SWinf\)_0$.

%%%%%%%%%%%%%%%%%%%%%%%%%%%%%%%%%
%%                             %%
%%  3.6                        %%
%%                             %%
%%%%%%%%%%%%%%%%%%%%%%%%%%%%%%%%%
%%%%%%%%%%%%%%%%%% supervariable %%%%%%%%%%%%%%%%%%%%%%%%%%%%%%%
\def\Fc{$[X,Y]\equiv XY-YX$, $\{X,Y\}\equiv XY+YX$.}
%%%%%%%%%%%
\no{\bf 3.6.}~
We let $\t$ be a Grassmann number, and identify
$$
\Smat{     P_0 & P_+ \cr    P_- & P_1      \cr}=
\Smat{\t\der\t & \t  \cr \der\t & \der\t\t \cr}.
$$
Then the multiplication $(f^A(D)P_A)\cdot(g^B(D)P_B)$ as superderivatives
corresponds to the multiplication
$$
\Fmat{f}{D}\cdot\Fmat{g}{D}
$$
as matrices.

The (anti--) commutation relations of superderivatives
(with central terms) are now easily obtained.
For example, setting $n>0$, we obtain{\foot{$^\dagger$}\Fc}
$$\eqalign{
\[W\(z^n f(D)P_a\),\,W\(z^m g(D)P_a\)\]
&=W\(\[z^n f(D)P_a,\,z^m g(D)P_a\]\)\cr
-(-1)^a\,&C \sum_{j=1}^n f(-j) g(n-j) \d_{n+m,0},\qquad (a=0,1),\cr
\mBRA W\(z^n f(D)P_\pm\),\,W\(z^m g(D)P_\mp\)\mKET
&=W\(\mBRA z^n f(D)P_\pm,\,z^m g(D)P_\mp\mKET\)\cr
&\mp\,C \sum_{j=1}^n f(-j) g(n-j) \d_{n+m,0}.\cr
}$$
Other (anti--) commutation relations have no central terms.

%%%%%%%%%%%%%%%%%%%%%%%%%%%%%%%%%%%%%%%%%%%%%%
%%                                          %%
%%  4. Quasifinite Representations          %%
%%                                          %%
%%%%%%%%%%%%%%%%%%%%%%%%%%%%%%%%%%%%%%%%%%%%%%%%%%%%%%%%%%%%%%%%%%%%%%
\bigbreak
\no{\bf 4. Quasifinite Representations}
\vskip2mm
%%%%%%%%%%%%%%%%%% highest weight module %%%%%%%%%%%%%%%%%%%%%%%%%%%%%%
\def\Everma{4.1}
\def\Ecp{4.2}
\def\Eai{4.3}
\def\Eaii{4.4}
\def\Ebi{4.5}
\def\Ebii{4.6}
\def\Elcma{4.7a}
\def\Elcmb{4.7b}
\def\Elcmc{4.7c}
\def\Elcmd{4.7d}
\def\ElcmA{4.8a}
\def\ElcmB{4.8b}
\def\ElcmC{4.8c}
\def\ElcmD{4.8d}
\def\Ecpk{4.9}
%%%%%%%%%%%%%%%%%%%%%%%%%%%%%%%%%
%%                             %%
%%  4.1                        %%
%%                             %%
%%%%%%%%%%%%%%%%%%%%%%%%%%%%%%%%%
\no{\bf 4.1.}~
Let $V(\la)$ be a highest weight module over $\SWinf$
with the highest weight $\la$.
The highest weight vector $\|\la\>\in V(\la)$ is characterized via the
principal gradation as %follows:
$\(\SWinf\)_\a\|\la\>=0$ for $\a\geq 1/2$ and
$\(\SWinf\)_0 \|\la\>\subset\bC\|\la\>$.
Explicitly, these conditions are written as
$$\eqalign{
W\(z^nf^A(D)P_A\)\|\la\>&=0\qquad\(n\ge 1\,;\,{}^\forall f^A(w)\in\bC[w]\),\cr
W\(   f  (D)P_+\)\|\la\>&=0\qquad\({}^\forall f(w)\in\bC[w]\),\cr
W\(     D^s P_a\)\|\la\>&=\D_a^{(s)}\,\|\la\>\qquad(s\ge 0\,;\,a=0,1)
}\eqno(\Everma)$$
for some functions $\D_a^{(s)}$ of $\la$.

It is convenient to introduce the generating functions
$\D_a(x)$ of highest weights $\D_a^{(s)}$ $(a=0,1)$:
$\D_a(x)\equiv-\sum_{s=0}^\infty\D_a^{(s)}x^s/s!$.
Note that they are formally given as the eigenvalues of the operators
$-W\(e^{xD}P_a\)$:
$$
W\(e^{xD}P_a\)\|\la\>=-\D_a(x)\|\la\>\qquad(a=0,1).
$$

%%%%%%%%%%%%%%%%%%%%%%%%%%%%%%%%%
%%                             %%
%%  4.2                        %%
%%                             %%
%%%%%%%%%%%%%%%%%%%%%%%%%%%%%%%%%
%%%%%%%%%%%%%%%%%%% quasifinite module %%%%%%%%%%%%%%%%%%%%%%%%%%%%%
\no{\bf 4.2.}~
Let $U(\la)$ be a subspace of $V(\la)$ which is obtained from
the highest weight state $\|\la\>$ by acting on it $\SWinf$ once:
$U(\la)=\SWinf\|\la\>$.
The principal gradation of $\SWinf$ naturally induces the labeling
of $U(\la)$:
$U(\la)=\bigoplus_{\a\ge 0}U_{-\a}(\la)$.

The $\SWinf$ module $V(\la)$ is called {\it quasifinite}
if $U_{-\a}(\la)$ is finite dimensional for each $\a\ge 0$.
This condition is equivalent to the statement that
$$
U^A_{-k}(\la)\equiv\set{W(z^{-k}f(D)P_A)\|\la\> }{f(w)\in\bC[w]}
$$
is finite dimensional for each $k\ge 0$ and $A=0,1,\pm$.
It is straightforward to see that the following subsets of $\bC[w]$
are all ideals of $\bC[w]$:
$$\eqalign{
I^-_0   &\equiv\set{f(w)\in\bC[w]}{W\(      f(D)P_-\)\|\la\>= 0},\cr
I^A_{-k}&\equiv\set{f(w)\in\bC[w]}{W\(z^{-k}f(D)P_A\)\|\la\>= 0}
\qquad (k\ge 1\,;\,\AA).}$$
Thus, if $U^A_{-k}(\la)$ is finite dimensional,
all of $I^-_0$ and $I^A_{-k}$ are different from $\{0\}$,
so that $I^-_0$ and $I^A_{-k}$ are generated by some monic polynomials
$a^-(w)$ and $b^A_k(w)$, respectively:
$I^-_0   =\(a^-(w)\)$ and $I^A_{-k}=\(b^A_k(w)\)$.
Conversely, if $I^-_0$ and $I^A_{-k}$ are generated by monic polynomials,
then $U^A_{-k}(\la)$ become finite dimensional since
$$
{\rm dim}\,U^A_{-k}(\la)
={\rm dim}\,\bC[w]/I^A_{-k}
={\rm deg}\,b^A_k(w)
< \infty.
$$
Thus, we have proved the following theorem:

\no{\bf\csc Theorem}.~{\it
The highest weight module $V(\la)$ of $\SWinf$ is quasifinite
if and only if the subsets
$I^-_0$ and $I^A_{-k}$ of $\bC[w]$ are generated by monic polynomials;
$$
I^-_0   =\(a^-(w)\)\qquad
I^A_{-k}=\(b^A_k(w)\)\qquad(k\ge 1\,;\,A=0,1,\pm).
\eqno(\Ecp)$$
}%%%%%%%%%%%%% end of the theorem
We will call $a^-(w)$, $b^A_k(w)$ the characteristic polynomials
for the highest weight module $V(\la)$.

For later convenience, we introduce the symbol
$$
B_k(w)\equiv b^A_k(w) P_A =\Fmat{b_k}{w},
$$
and further denote
$b^A(w)\equiv b^A_1(w)$, $B(w)\equiv B_1(w)$.
In the following discussions, we will see that
$a^-(w)$ and $b^+(w)$ play the central role in
the quasifinite representations of $\SWinf$.

%%%%%%%%%%%%%%%%%%%%%%%%%%%%%%%%%
%%                             %%
%%  4.3                        %%
%%                             %%
%%%%%%%%%%%%%%%%%%%%%%%%%%%%%%%%%
%%%%%%%%%%%%%%%%%%%% level 0 conditions %%%%%%%%%%%%%%%%%%%%%%%%%%
\def\Fapi{Another proof of Theorems 4.3 and 4.4 is given in Appendix A,
resorting to the embedding of $\SWinf$ into $\Glinf$.
}%%%%%%%%%%%%
\no{\bf 4.3.}~
{\bf\csc Theorem}.~{\it
Characteristic polynomials $a^-(w)$, $b^A(w)$ $(\AA)$
are related to each other in the following manner:
$$\eqalign{
&a^-(w)           ~~|~~b^0(w),\cr
&         a^-(w-1)~~|~~b^1(w),\cr
&a^-(w),~~a^-(w-1)~~|~~b^-(w),\cr
}\eqno(\Eai)$$
and
$$\eqalign{
&b^+(w)~~|~~        b^0(w),~~      b^1(w),\cr
&b^0(w)~~|~~a^-(w  )b^+(w),~~      b^-(w),\cr
&b^1(w)~~|~~a^-(w-1)b^+(w),~~      b^-(w),\cr
&b^-(w)~~|~~a^-(w-1)b^0(w),~~a^-(w)b^1(w).\cr
}\eqno(\Eaii)$$
Here $f_1(w),\cdots,f_r(w)\,|\,g_1(w),\cdots,g_s(w)$
implies that any $f_i(w)$ divides all $g_j(w)$'s.
}%%%%%%%%%% end of the theorem %%%%%%%%%%%%

\no{\it Proof }.{\foot{$^\dagger$}{\Fapi}}
We start from the identity
$$
\[W\(z\Emat\),W\(z^{-1}\aBmat{}\)\mKET\|\la\> = 0
$$
which holds for arbitrary constants
$\a$, $\b$, $\g$, $\d$ and $\e^A$ $(\AA)$.
Suitably choosing these constants,
we can derive the following equations:
$$\eqalign{
&W\(\Smat{0 &0 \cr b^0(D)   &0 \cr}\)\|\la\>=0,\cr
&W\(\Smat{0 &0 \cr b^1(D+1) &0 \cr}\)\|\la\>=0,\cr
&W\(\Smat{0 &0 \cr b^-(D)   &0 \cr}\)\|\la\>=
 W\(\Smat{0 &0 \cr b^-(D+1) &0 \cr}\)\|\la\>=0,\cr
}$$
which assert the first statement, eq.\ (\Eai).
The second statement, eq.\ (\Eaii), can be similarly proved,
by using the identity
$$
\[W\(\Smat{0&\e^+\cr\e^-a^-(D)&0\cr}\),W\(z^{-1}\aBmat{}\)\mKET\|\la\> = 0,
$$
and taking a suitable choice of the constants
$\a$, $\b$, $\g$, $\d$ and $\e^\pm$.
\qed

%%%%%%%%%%%%%%%%%%%%%%%%%%%%%%%%%
%%                             %%
%%  4.4                        %%
%%                             %%
%%%%%%%%%%%%%%%%%%%%%%%%%%%%%%%%%
%%%%%%%%%%%%%%%%%%%% level r conditions %%%%%%%%%%%%%%%%%%%%%%%%%%%%%%
\no{\bf 4.4.}~
{\bf\csc Theorem}.~{\it
Characteristic polynomials $b^A_k(w)$ for $k\geq 1$
are related to each other in the following manner:
$$\eqalign{
b^+_k(w),~~b^0_k(w-1),~~b^1_k(w),~~b^+_k(w-1)~~&|~~b^+_{k+1}(w),\cr
b^0_k(w),~~b^0_k(w-1),~~b^-_k(w),~~b^+_k(w-1)~~&|~~b^0_{k+1}(w),\cr
b^1_k(w),~~b^1_k(w-1),~~b^+_k(w),~~b^-_k(w-1)~~&|~~b^1_{k+1}(w),\cr
b^-_k(w),~~b^1_k(w-1),~~b^0_k(w),~~b^-_k(w-1)~~&|~~b^-_{k+1}(w),\cr
}\eqno(\Ebi)$$
and
$$\eqalign{
b^+_{k+\l}(w)~~&|~~b^0_k(w-\l)b^+_\l(w),~~b^+_k(w-\l)b^1_\l(w),\cr
b^0_{k+\l}(w)~~&|~~b^0_k(w-\l)b^0_\l(w),~~b^+_k(w-\l)b^-_\l(w),\cr
b^1_{k+\l}(w)~~&|~~b^-_k(w-\l)b^+_\l(w),~~b^1_k(w-\l)b^1_\l(w),\cr
b^-_{k+\l}(w)~~&|~~b^-_k(w-\l)b^0_\l(w),~~b^1_k(w-\l)b^-_\l(w).\cr
}\eqno(\Ebii)$$
}%%%%%%%%%% end of the theorem %%%%%%%%%%%%

\no{\it Proof }.
The first statement, eq.\ (\Ebi), is obtained by
combining the following two identities:
$$\eqalign{
&\[W\(z\Emat\),W\(z^{-k-1}\aBmat{k+1}\)\mKET\|\la\>=0,\cr
&\[W\(z\Emat\),W\(z^{-k-1}D\aBmat{k+1}\)\mKET\|\la\>=0
}$$
which hold for arbitrary constants
$\a$, $\b$, $\g$, $\d$ and $\e^A$ $(\AA)$.
The second statement, eq.\ (\Ebii), is obtained by
looking at the identities
$$\eqalign{
&\[W\(z^{-k}\Fmat{b_k}{D}\),W\(z^{-\l} \aBmat{\l}\)\mKET\|\la\>=0,\cr
&\[W\(z^{-k}\Fmat{b_k}{D}\),W\(z^{-\l}D\aBmat{\l}\)\mKET\|\la\>=0.
}$$
\rightline{\qed}

\no
Note that if we set
$b^-_0(w)\equiv a^-(w)$, $b^+_0(w)\equiv 1$ and
$b^0_0(w)\equiv b^1_0(w)\equiv 0$,
then the Theorem 4.4 reduces to the Theorem 4.3
with some suitable choices of $k$ and $\l$.

%%%%%%%%%%%%%%%%%%%%%%%%%%%%%%%%%
%%                             %%
%%  4.5                        %%
%%                             %%
%%%%%%%%%%%%%%%%%%%%%%%%%%%%%%%%%
%%%%%%%%%%%%%%%% solution for characteristic polinomials %%%%%%%%%%%%%%%%
\def\Fapii{This equation is derived in a simpler way in Appendix.
}%%%%%%%%%%%
\no{\bf 4.5.}~
Iteratively using Theorems 4.3 and 4.4, we obtain the following Corollary:

\no{\bf\csc Corollary}.~{\it
Characteristic polynomials $b^A_k(w)$ for $k\ge 1$
are related to the polynomials $a^-(w)$ and $b^+(w)$ as
$$\eqalign{
\l.c.m.\(b^+(w),a^-(w-1),b^+(w-1),a^-(w-2),\cdots,b^+(w-k+1)\)
{}~&|~b^+_k(w),\cr
\l.c.m.\(a^-(w),b^+(w),a^-(w-1),b^+(w-1),\cdots,b^+(w-k+1)\)
{}~&|~b^0_k(w),\cr
\l.c.m.\(b^+(w),a^-(w-1),b^+(w-1),a^-(w-2),\cdots,a^-(w-k)\)
{}~&|~b^1_k(w),\cr
\l.c.m.\(a^-(w),b^+(w),a^-(w-1),b^+(w-1),\cdots,a^-(w-k)\)
{}~&|~b^-_k(w),\cr
}\quad\eqalign{
(\Elcma)\cr(\Elcmb)\cr(\Elcmc)\cr(\Elcmd)\cr
}$$
and
$$\eqalign{
b^+_k(w)~~&|~~
b^+(w)\,a^-(w-1)\,b^+(w-1)\,a^-(w-2)\,\cdots\,b^+(w-k+1),\cr
b^0_k(w)~~&|~~
a^-(w)\,b^+(w)\,a^-(w-1)\,b^+(w-1)\,\cdots\,b^+(w-k+1),\cr
b^1_k(w)~~&|~~
b^+(w)\,a^-(w-1)\,b^+(w-1)\,a^-(w-2)\,\cdots\,a^-(w-k),\cr
b^-_k(w)~~&|~~
a^-(w)\,b^+(w)\,a^-(w-1)\,b^+(w-1)\,\cdots\,a^-(w-k).\cr
}\eqno\eqalign{
(\ElcmA)\cr(\ElcmB)\cr(\ElcmC)\cr(\ElcmD)\cr
}$$
}%%%%%%%%%% end of the theorem %%%%%%%%%%%%

Let $a^-(w)=\prod_{i=1}^{N_-}(w-\la^-_i)$ and
$b^+(w)=\prod_{j=1}^{N_+}(w-\la^+_j)$.
If difference of any two distinct elements
of the set $\{\la^-_i\}\cup\{\la^+_j\}$ is not an integer,
then $a^-(w)$, $b^+(w)$, $a^-(w-1)$, $b^+(w-1)$, $\cdots$
are all mutually prime.
In this case, the characteristic plynomials $b^A_k(w)$ $(k\ge1,\,\AA)$ are
uniquely determined due to the above corollary
as follows:{\foot{$^\dagger$}{\Fapii}}
$$\eqalign{
B_k(w)
&=\(\prod_{\l=0}^{k-1} b^+(w-\l)\)\Smat{
\prod_{\l=0}^{k-1}a^-(w-\l) &\prod_{\l=1}^{k-1}a^-(w-\l)\cr
\prod_{\l=0}^{k  }a^-(w-\l) &\prod_{\l=1}^{k  }a^-(w-\l)\cr}\cr
&=\(\prod_{\l=1}^{k-1} a^-(w-\l)b^+(w-\l)\)
\Smat{a^-(w)        b^+(w)  &        b^+(w)\cr
      a^-(w)a^-(w-k)b^+(w)  &a^-(w-k)b^+(w)\cr}\cr
&={1\over 2^{k-1}}B(w-k+1)\cdots B(w-1) B(w).
}\eqno(\Ecpk)$$

%%%%%%%%%%%%%%%%%%%%%%%%%%%%%%%%%%%%%%%%%%%%%%%%%%%%
%%                                                %%
%%  5. Differential Equations for Highest Weights %%
%%                                                %%
%%%%%%%%%%%%%%%%%%%%%%%%%%%%%%%%%%%%%%%%%%%%%%%%%%%%%%%%%%%%%%%%%%%%%
\bigbreak
\no{\bf 5. Differential Equations for Highest Weights}
\vskip2mm
%%%%%%%%%%%%%%%%%%%%%%%%%%%%%%%%%%%%%%%%%%%%%%%%%%%%%%%%%%%%%%%%%%%%%%
\def\Edi{5.1}
\def\Ediia{5.2a}
\def\Ediib{5.2b}
\def\Ediic{5.2c}
\def\Ediid{5.2d}
\def\Ediii{5.3}
\def\Ediv{5.4}
\def\EDi{5.5}
\def\EDii{5.6}
\def\EDa{5.7a}
\def\EDb{5.7b}
\def\EDc{5.7c}
\def\EDd{5.7d}
%%%%%%%%%%%%%%%%%%%%%%%%%%%%%%%%%
%%                             %%
%%  5.1                        %%
%%                             %%
%%%%%%%%%%%%%%%%%%%%%%%%%%%%%%%%%
\no{\bf 5.1.}~
The structure of characteristic polynomials
automatically determines that of highest weights.
In the following subsections,
we derive the differential equations for $\D_a(x)$ $(a=0,1)$.
Recall that $W\(e^{xD} P_a\)\|\la\>=-\D_a(x)\|\la\>$.

%%%%%%%%%%%%%%%%%%%%%%%%%%%%%%%%%
%%                             %%
%%  5.2                        %%
%%                             %%
%%%%%%%%%%%%%%%%%%%%%%%%%%%%%%%%%
%%%%%%%%%%%%%%%%% level 0 equation %%%%%%%%%%%%%%%%%%%%%%%%%%%
\no{\bf 5.2.}~
We first note that for arbitrary functions $f(w)\in\bC[w]$,
the following equation holds:
$$
\mBRA
W\(\Smat{0 & f(D) \cr 0 & 0 \cr}\),
W\(\Smat{0 & 0 \cr a^-(D) & 0 \cr}\)\mKET\|\la\>=0.
$$
The left--hand side can be rewritten as $W\(f(D)a^-(D)(P_0+P_1)\)\|\la\>$,
and thus, by setting  $f(D)=\exp(xD)$, we obtain
$$
a^-\({d\over dx}\) \[\D_0(x)+\D_1(x)\]=0.
\eqno(\Edi)$$

%%%%%%%%%%%%%%%%%%%%%%%%%%%%%%%%%
%%                             %%
%%  5.3                        %%
%%                             %%
%%%%%%%%%%%%%%%%%%%%%%%%%%%%%%%%%
%%%%%%%%%%%%%%%%%% level 1 equations %%%%%%%%%%%%%%%%%%%%%%%%%%
\no{\bf 5.3.}~
We then use the identity
$\[W\(zG(D+1)\),W\(z^{-1}B(D)\)\]\|\la\>=0$
which holds for arbitrary element $G(w)=g^A(w)P_A\in\cA$.
If we set $g^A(D)=\a^A\exp(xD)$
and pick up the coefficient of $\a^A$ $(\AA)$,
we obtain the following set of differential equations:
$$\eqalign{
b^+&\({d\over dx}\) \[    e^x  \D_0(x) +          \D_1(x) - C \]=0,\cr
b^0&\({d\over dx}\) \[\(1-e^x\)\D_0(x)                    + C \]=0,\cr
b^1&\({d\over dx}\) \[                   \(1-e^x\)\D_1(x) - C \]=0,\cr
b^-&\({d\over dx}\) \[         \D_0(x) +     e^x  \D_1(x) + C \]=0.\cr
}\eqno\eqalign{
{{}^{}\atop{}^{}}(\Ediia)\cr
{{}^{}\atop{}^{}}(\Ediib)\cr
{{}^{}\atop{}^{}}(\Ediic)\cr
{{}^{}\atop{}^{}}(\Ediid)\cr
}$$

Surprisingly, all of these four equations reduce to the first one
if we use eq.\ (\Edi).
To prove this, we first notice that eq.\ (\Edi) can be rewritten as
$$
a^-\({d\over dx}-1\) \[e^x\D_0(x)+e^x\D_1(x)\]=0.
\eqno(\Ediii)$$
Since eq.\ (\Eaii) implies that
$b^0(w)$, $b^1(w)$ and $b^-(w)$ are all divided by $b^+(w)$,
we can replace $b^+\(d/dx\)$ in eq.\ (\Ediia)
by $b^A\(d/dx\)$ $(\AA)$:
$$
b^A\({d\over dx}\) \[e^x\D_0(x)+\D_1(x)-C\]=0\qquad(\AA).
$$
As for $A=0$, $\D_1(x)$ can be replaced by $-\D_0(x)$,
since $b^0(w)$ is divided by $a^-(w)$.
As for $A=1$, $e^x\D_0(x)$ can be replaced by $-e^x\D_1(x)$,
since $b^1(w)$ is divided by $a^-(w-1)$
and so we can use eq.\ (\Ediii).
Finally as for $A=-$, $e^x\D_0(x)$ and $\D_1(x)$
can be replaced by $-e^x\D_1(x)$ and $-\D_0(x)$, respectively,
since $b^-(w)$ is divided by both of $a^-(w-1)$ and $a^-(w)$.

%%%%%%%%%%%%%%%%%%% differential equations %%%%%%%%%%%%%%%%%%%

We summarize the results obtained above in the following theorem:

\no{\bf\csc Theorem}.~{\it
The generating functions $\D_a(x)$ $(a=0,1)$ of highest weights
satisfy the following differential equations:
$$\eqalign{
&a^-\({d\over dx}\) \[   \D_0(x) + \D_1(x) \]=0,\cr
&b^+\({d\over dx}\) \[e^x\D_0(x) + \D_1(x) - C \]=0.
}\eqno(\Ediv)$$
}%%%%%%%%%% end of the theorem %%%%%%%%%%%%
%\no Note that the second equation can be rewritten into more symmetric form;
%$$b^+\({d\over dx}\) \[(e^x-1)(\D_0(x) - \D_1(x)) + 2C \]=0.$$

%%%%%%%%%%%%%%%%%%%%%%%%%%%%%%%%%
%%                             %%
%%  5.4                        %%
%%                             %%
%%%%%%%%%%%%%%%%%%%%%%%%%%%%%%%%%
%%%%%%%%% Solutions %%%%%%%%%%%%%%%%%%%%%%%%%%%%
\no{\bf 5.4.}~
We assume that polynomials $a^-(w)$, $b^+(w)$ have the following form:
$$
a^-(w) =\prod_{i=1}^M(w-\mu_i)^{m_i},\qquad
b^+(w) =\prod_{j=1}^N(w-\nu_j)^{n_j},
\eqno(\EDi)$$
where $\mu_i\neq\mu_{i'}$ if $i\neq i'$,
and   $\nu_j\neq\nu_{j'}$ if $j\neq j'$.
Then the differential equations (\Ediv) may be solved as
$$\eqalign{
\D_0(x)+\D_1(x)&=\sum_{i=1}^M p_i(x) e^{\mu_i x},\cr
e^x \D_0(x)+\D_1(x)-C&=-\sum_{j=1}^N q_j(x) e^{\nu_j x}.\cr
}$$
Here $p_i(x)$ and $q_j(x)$ are, respectively,
degree $m_i-1$ and $n_j-1$ polynomials of $x$.
Since these equations can be rewritten as
$$\eqalign{
\D_0(x)&=-{\sum_{i=1}^M p_i(x) e^{\mu_i x}
          +\sum_{j=1}^N q_j(x) e^{\nu_j x} - C \over e^x-1}, \cr
\D_1(x)&=+{\sum_{i=1}^M p_i(x) e^{(\mu_i+1) x}
          +\sum_{j=1}^N q_j(x) e^{ \nu_j    x} - C \over e^x-1},
}\eqno(\EDii)$$
we obtain four typical representations,
$$\eqalign{
\D_0(x)&=-\,C\,{{e^{\la x}-1}\over{e^x-1}},\cr
\D_0(x)&=-\,{{q_k x^ke^{\la x}}\over{e^x-1}},\cr
\D_0(x)&=-\,C\,{{e^{\la x}-1}\over{e^x-1}},\cr
\D_0(x)&=-\,{{p_k x^k e^{\la x}}\over{e^x-1}},\cr
}\qquad\eqalign{
\D_1(x)&=+\,C\,{{e^{\la x}-1}\over{e^x-1}},\cr
\D_1(x)&=+\,{{q_k x^ke^{\la x}}\over{e^x-1}},\cr
\D_1(x)&=+\,C\,{{e^{(\la+1) x}-1}\over{e^x-1}},\cr
\D_1(x)&=+\,{{p_k x^k e^{(\la+1) x}}\over{e^x-1}},\cr
}\eqno\eqalign{
{{}^{}\atop{}^{}}(\EDa)\cr
{{}^{}\atop{}^{}}(\EDb)\cr
{{}^{}\atop{}^{}}(\EDc)\cr
{{}^{}\atop{}^{}}(\EDd)\cr
}$$
with $k\in\bZ_{>0}$.
Here eq.\ (\EDa) corresponds to the case where $a^-(w)=1$ and $b^+(w)=w-\la$,
and eq.\ (\EDc) to the case where $a^-(w)=w-\la$ and $b^+(w)=1$.
Eq.\ (\EDb) corresponds to the special case where
$a^-(w)=1$, $b^+(w)=(w-\la)^{k+1}$ and $C=0$,
while eq.\ (\EDd) to the special case where
$a^-(w)=(w-\la)^{k+1}$, $b^+(w)=1$ and $C=0$ [M].
First two solutions describe the system
having no degeneracy in the vacuum ($a^-(w)=1$).
On the other hand, in the last two representations,
we have several states at level 0.
%In section 7, we show their explicit realizations in terms of free fields.

%%%%%%%%%%%%%%%%%%%%%%%%%%%%%%%%%%%%%%%%%%%%%%
%%                                          %%
%%  6. Spectral Flow                        %%
%%                                          %%
%%%%%%%%%%%%%%%%%%%%%%%%%%%%%%%%%%%%%%%%%%%%%%%%%%%%%%%%%%%%%%%%%%%%%%%
\bigbreak
\no{\bf 6. Spectral Flow}
\vskip2mm
%%%%%%%%%%%%%%%%%%%% Spectal Flow %%%%%%%%%%%%%%%%%%%%%%%%%%%%%%%%%%%%%%
\def\Esf{6.1}
\def\Esfi{6.2}
\def\Esfiib{6.3}
\def\Esfiiib{6.4}
\def\Esfiia{6.5}
\def\Esfiiia{6.6}
%%%%%%%%%%%%%%%%%%%%%%%%%%%%%%%%%
%%                             %%
%%  6.1                        %%
%%                             %%
%%%%%%%%%%%%%%%%%%%%%%%%%%%%%%%%%
\no {\bf 6.1.}~
Since $\SWinf$ contains two $u(1)$ Kac-Moody algebras as subalgebras,
$\SWinf$ has a two--parameter family of automorphisms
which we will call the {\it spectral flow}.

\no{\bf\csc Theorem}.~{\it
There exist the following automorphisms $W(\cdot)\mapsto W'(\cdot)$:
$$\eqalign{
  W'(z^n e^{xD} P_a)
  &=W(z^n e^{x(D+\la^a)}P_a)\pm C {e^{\la^ax}-1 \over e^x-1} \d_{n0},
  \quad a=\{ {}^0_1, \cr
  W'(z^n e^{xD} P_{\pm})
  &=W(z^{n \pm (\la^1-\la^0)} e^{x(D+\la^a)}P_{\pm}),
  \quad\qquad\qquad a=\{ {}^1_0,
}\eqno{(\Esf)}$$
with arbitrary parameters $\la^a$ $(a=0,1)$.
}%%%%%%%%%%%%%%%%%%% end of theorem

\no{\it Proof }.
One can easily show that this new generators $W'(\cdot)$ satisfy
the same (anti--) commutation relations as those for the original ones
$W(\cdot)$, eq.\ (\Ecrii). \qed

%%%%%%%%%%%%%%%%%%% N=2 SCA %%%%%%%%%%%%%%%%%%%%%%%%%%%%%%%%%%%
%%%%%%%%%%%%%%%%%%%%%%%%%%%%%%%%%
%%                             %%
%%  6.2                        %%
%%                             %%
%%%%%%%%%%%%%%%%%%%%%%%%%%%%%%%%%
\no {\bf 6.2.}~
Under the spectral flow, the highest weight state may change
although the representation space as a set is kept invariant.
We illustrate this phenomena by taking the $N=2$ superconformal algebra
[SS] as an example (see figure 1).

%%%%%%%%%%%%%%%%%%%%%%  Fig. %%%%%%%%%%%%%%%%%%%%
\input epsf.tex
\vbox{
\centerline{\epsfbox{swfig1.eps}}
\centerline{
{\bf Figure 1.} Spectral Flow for the $N=2$ Superconformal Algebra}
}%%%%%%%%%%%  End of Fig.  %%%%%%%%%%%%%

The generators of the $N=2$ superconformal algebra
consist of $J_n$ ($U(1)$--current),
$L_n$ (energy--momentum tensor) and $G^{\pm}_r$ (supercurrents),
and satisfy the following (anti--) commutation relation:
$$\eqalign{&\eqalign{
\[L_n,L_m    \]&=(n-m)L_{n+m}+{c\over 12}(n^3-n)\d_{n+m,0},\cr
\[L_n,G_r^\pm\]&=\({n\over 2}-r\)G_{n+r}^\pm,\cr
\mBRA G_r^+  ,G_s^-  \mKET
&=2L_{r+s}+(r-s)J_{r+s}+{c\over 3}\(r^2-{1\over 4}\)\d_{r+s,0},\cr
}\cr &\eqalign{
\mBRA G_r^\pm,G_s^\pm\mKET&=0,\cr
\[J_n,J_m    \]&={c\over 3}n\d_{n+m,0},\cr
}\qquad\eqalign{
\[L_n,J_m    \]&=-m J_{n+m},\cr
\[J_n,G_r^\pm\]&=\pm G_{n+r}^\pm.\cr
}}$$
Here $n \in \bZ$, and
$r\in\bZ+1/2$ for Neveu--Schwarz ($NS$) sector or
$r\in\bZ    $ for Ramond         ($R $) sector.
The highest weight state $|q,h\>$ is characterized by
$$\eqalign{
  J_n, L_n, G^+_r, G^-_s |q,h\> &= 0 \quad (n>0, r \geq 0, s>0), \cr
  J_0 |q,h\> &= q |q,h\>, \cr
  L_0 |q,h\> &= h |q,h\>.
}$$

%%%%%%%%%%%%%%%% spectral fow for N=2 %%%%%%%%%%%%%%%%%%%%%%%%%%%%
This algebra is invariant under the following transformation with
arbitrary parameter $\la$:
$$\eqalign{
  J'_n &= J_n +{c \over 3} \la \d_{n0}, \cr
  L'_n &= L_n +\la J_n+{c \over 6} \la^2 \d_{n0}, \cr
  G'^{\pm}_r &= G^{\pm}_{r \pm \la}.
}$$
When $\la$ is an integer (or half--odd integer),
the spectral flow maps
$NS$ sector to $NS$ sector and  $R$ sector to $R$ sector
(or $NS$ to $R$, $R$ to $NS$).

We first consider the case of $R$ to $R$ with $\la =1$.
For $h \neq c/24$, $|q,h\>$ is no longer the highest weight
state with respect to the new generators, because
$G'^-_1 |h,q\> = G^-_0 |q,h\>$ does not vanish.
However, the new state $G^-_0 |q,h\>$
satisfies the highest weight condition,
and may be identified with the new highest weight state $|q',h'\>'$.
Here, the new $u(1)$ charge $q'$ and the new conformal weight $h'$ are
$$
q'=q-1+c/3\qquad h'=h+q-1+c/6,
$$
because
$J'_0|q',h'\>'=(    J_0+c/3)G^-_0|q,h\> =(  q-1 +c/3) G^-_0|q,h\>$ and
$L'_0|q',h'\>'=(L_0+J_0+c/6)G^-_0|q,h\> =(h+q-1 +c/6) G^-_0|q,h\>$.
For $h=c/24$, the new highest weight state is given by
$|q',h'\>' = |q,h\>$ with $q'=q+c/3$ and $h'=h+q+c/6$.

Similarly, in the case of $R$ to $R$ with $\la=-1$,
the new highest weight state for $h-q+c/8 \neq 0$ is given by
$|q',h'\>'= G^+_{-1} |q,h\>$ with $q'=q+1-c/3$ and $h'=h-q+c/6$.

%%%%%%%%%%%%%%%%%%% The new weight %%%%%%%%%%%%%%%%%%%%%%%%
%%%%%%%%%%%%%%%%%%%%%%%%%%%%%%%%%
%%                             %%
%%  6.3                        %%
%%                             %%
%%%%%%%%%%%%%%%%%%%%%%%%%%%%%%%%%
\def\Fnull{
When eq.\ (\Esfiia) is a null state,
we must replace the upper bound of the product by a smaller number.
}%%%%%%%%%%%%%%%%%
\no {\bf 6.3.}~
Let us go back to $\SWinf$.
We would like to derive the modification of the weights
and the characteristic polynomials under the spectral flow.
We restrict ourselves to the case $\la^1-\la^0 \in \bZ$.
Thus, it is sufficient to consider three cases $\la^1-\la^0=0,\pm1$,
because, for example, the flow with $\la^1-\la^0=2$ is
obtained by taking twice the flow with $\la^1-\la^0=1$.
We have the following Theorem:

\no{\bf\csc Theorem}.~{\it
Under the spectral flow, the new weights, $\D'_a(x)$,
and the new characteristic polynomials, $a'^-(w)$ and $b'^+(w)$,
are given as follows, for generic values of $C$ and $\D_a(x)$:

\no{\rm (\romannumeral1)} If $\la^1-\la^0=0$, then
$$
\D'_a(x)=e^{\la^a x}\D_a(x)\mp C{e^{\la^a x}-1\over e^x-1},\quad a=\{ {}^0_1,
\eqno{(\Esfi)}$$
and $a'^-(w)=a^-(w-\la^0)$, $b'^+(w)=b^+(w-\la^1)$.

\no{\rm (\romannumeral2)} If $\la^1-\la^0=1$, then
$$
  \D'_a(x) =e^{\la^a x} \D_a(x)\mp C {e^{\la^a x}-1 \over e^x-1}
  \pm \sum_{i=1}^{N_-} e^{(\la^-_i +\la^a) x},\quad  a=\{ {}^0_1,
\eqno{(\Esfiib)}$$
and $a'^-(w)=b^-(w-\la^0)$, $b'^+(w)=a^-(w-\la^1)$.

\no{\rm (\romannumeral3)} If $\la^1-\la^0=-1$, then
$$
  \D'_a(x)=e^{\la^a x} \D_a(x)\mp C {e^{\la^a x}-1 \over e^x-1}
  \mp \sum_{j=1}^{N_+} e^{(\la^+_j +\la^1) x},\quad  a=\{ {}^0_1,
\eqno{(\Esfiiib)}$$
and $a'^-(w)=b^+(w-\la^0+1)$, $b'^+(w)=b^+_2(w-\la^1)$.
}%%%%%%%%%%%%%

\no
Eq. (\Esfi) is identical with the formula in the bosonic case [AFMO1].

\no{\it Proof }.

%%%%%%%%%%%%%%%%%%%% case 1 %%%%%%%%%%%%%%%%%%%%%%%%
\no
(\romannumeral1) $\la^1-\la^0=0$. \par
\no
The highest weight state $|\la\>$
with respect to the original generators $W$ is also
the highest weight state $|\la'\>'$ with respect to the new ones $W'$,
$|\la'\>'=|\la\>$.
Hence, the new weights, $\D'_a(x)$, are given by eq. (\Esfi).
Since $\|\la'\>'=\|\la\>$, we also have the following equation:
$$\eqalign{
W'(      a^-(D-\la^0)P_-)\|\la'\>'&=W(      a^-(D)P_-)\|\la'\>'=0,\cr
W'(z^{-1}b^+(D-\la^1)P_+)\|\la'\>'&=W(z^{-1}b^+(D)P_+)\|\la'\>'=0.
}$$
Therefore, the new characteristic polynomials are given by
$a'^-(w)=a^-(w-\la^0)$ and $b'^+(w)=b^+(w-\la^1)$.

%%%%%%%%%%%%%%%%%%%%%% case 2 %%%%%%%%%%%%%%%%%%%%%%%%%%
\no
(\romannumeral2) $\la^1-\la^0=1$. \par
\no
In this case $|\la\>$ is not the highest weight state
with respect to the new generators $W'$.
In generic situations{\foot{$^\dagger$}{\Fnull}},
the new highest weight state is given by
$$
  |\la'\>'=\prod_{k=0}^{N_- -1} W(D^k P_-) |\la\>,
\eqno{(\Esfiia)}$$
if $a^-(w)=\prod_{i=1}^{N_-} (w-\la^-_i)$.
To prove it, we first remark that
$W'(z^{n+1} f(D) P_0)$, \BR $W'(z^{n+1} f(D) P_1)$,
$W'(z^n f(D) P_+)$ and $W'(z^{n+2} f(D) P_-)$ with $n\geq 0$
annihilate $|\la'\>'$ in a trivial way.
On the other hand,
$W'(z f(D) P_-)$ annihilates $|\la'\>'$ since
the state $W'(z f(D) P_-) |\la'\>' $
can be rewritten in the following form:
$$\eqalign{
  W'(z f(D) P_-) |\la'\>' & =
  (-1)^{N_-} \prod_{k=0}^{N_- -1} W(D^k P_-) \cdot
  W(f(D + \la^0) P_-) |\la\> \cr
  & =(-1)^{N_-} \prod_{k=0}^{N_- -1} W(D^k P_-) \cdot
  \sum_{\l=0}^{N_- -1} c_{\l} W(D^{\l} P_-) |\la\>,
}$$
where $c_{\l}$ are some constants.
In this expression, we have replaced $f(D+\la^0)$
by a polynomial with degree less than $N_-$,
making use of the quasifinite condition $W(a^-(D) P_-) |\la\>$ $=0$.
This state vanishes because $W(D^k P_-)^2 = 0$.

The weights of this new highest weight state $|\la'\>'$
are calculated as follows.
First, we note the following equation:
$$\eqalign{
  -W' & (e^{xD} P_a) |\la'\>' \cr
  = &\prod_{k=0}^{N_- -1} W(D^k P_-) \cdot
  \( -W(e^{(D+\la^a) x} P_a)\mp C {e^{\la^a x}-1 \over e^x-1}\)|\la\> \cr
  & \pm \sum_{k=0}^{N_- -1}
  \prod_{k_2=k+1}^{N_- -1} W(D^{k_2} P_-) \cdot
  W(e^{x(D+\la^a)} D^k P_-)\cdot\prod_{k_1=0}^{k-1} W(D^{k_1} P_-)|\la\> \cr
 =&\( e^{\la^a x} \D_a(x)\mp C {e^{\la^a x}-1 \over e^x-1}\)|\la'\>' \cr
  & \pm \sum_{k=0}^{N_- -1}
  \prod_{k_2=k+1}^{N_- -1} W(D^{k_2} P_-) \cdot
  e^{\la^a x} W(r^-_k(D,x) P_-) \cdot
  \prod_{k_1=0}^{k-1} W(D^{k_1} P_-)|\la\>.
}$$
Here we first moved $W(e^{x(D+\la^A)} D^k P_-)$ to the right,
and then, after reducing the degree in $D$ using the quasifinite
condition $W(a^-(D) P_-) |\la\> = 0$,
we substituted it into the original position.
The function $r^-_k(D,x)=\sum_{\l=0}^{N_- -1} r^-_{k,\l}(x) D^{\l}$
is defined as a remainder of $e^{xD} D^k$ by $a^-(D)$:
$e^{xD} D^k=a^-(D) q^-_k(D,x) +r^-_k(D,x)$.
Note that only $D^k$ term in $r^-_k(D,x)$ contributes because
$W(D^{\l} P_-)^2 = 0$. Furthermore,
we can also show that $r^-_{k,k}(x) = ({d \over dx})^k r^-_{0,k}(x)$
and $\sum_{k=0}^{N_- -1} ({d \over dx})^k r^-_{0,k}(x) =
\sum_{i=1}^{N_-} e^{\la^-_i x}$. We thus obtain eq.\ (\Esfiib).

Moreover, one may show that
$W(z^{-1}b^-(D)P_-)\|\la'\>' =0$ and
$W(      a^-(D)P_+)\|\la'\>' =0$ if and only if
$W(z^{-1}b^-(D)P_-)\|\la\> =0$ and
$a^-\({d \over dx}\)[\D_0(x)+\D_1(x)] = 0$, respectivly.
This can be proved as follows:
First, since $\mBRA W(z^m f(D)P_-),W(z^n g(D)P_-)\mKET=0$,
$$
W(z^{-1}b^-(D)P_-)\|\la'\>'
=(-1)^{N_-}\prod_{k=0}^{N_- -1} W(D^k P_-) W(z^{-1}b^-(D)P_-)\|\la\>.
$$
Second, since $\[W\(f(D)(P_0+P_1)\),W(g(D)P_-)\]=0$,
$$\eqalign{
W(&a^-(D)P_+)\|\la'\>'
=\sum_{\l=0}^{N_- -1}(-1)^{N_- -1-\l}
\prod_{k=0\atop k\neq\l}^{N_- -1}W(D^k P_-)
\cdot W\(a^-(D)D^\l (P_0+P_1)\)\|\la\> \cr
&=\sum_{\l=0}^{N_- -1}(-1)^{N_- -\l}
\prod_{k=0\atop k\neq\l}^{N_- -1}W(D^k P_-)\|\la\>
\, a^-\({d\over dx}\)\({d\over dx}\)^\l\[\D_0(x)+\D_1(x)\]\Big|_{x=0}.\cr
}$$
Hence,
$$\eqalign{
W'(      b^-(D-\la^0)P_-)\|\la'\>'=W(z^{-1}b^-(D)P_-)\|\la'\>'=0,\cr
W'(z^{-1}a^-(D-\la^1)P_+)\|\la'\>'=W(      a^-(D)P_+)\|\la'\>'=0.
}$$
Therefore, the new characteristic polynomials are given by
$a'^-(w)=b^-(w-\la^0)$ and $b'^+(w)=a^-(w-\la^1)$.

%%%%%%%%%%%%%%%%%% case 3 %%%%%%%%%%%%%%%%%%%%%%
\no
(\romannumeral3) $\la^1-\la^0=-1$. \par
\no
Similarly to the case (\romannumeral2),
in generic situations the new highest weight state is given by
$$
  |\la'\>'=\prod_{k=0}^{N_+ -1} W(z^{-1} D^k P_+) |\la\>,
\eqno{(\Esfiiia)}
$$
if $b^+(w)=\prod_{j=1}^{N_+} (w-\la^+_j)$.
The weights of this new state are also similarly calculated.

Moreover, one may show that
$W(z     b^+(D+1)P_-)\|\la'\>' =0$ and
$W(z^{-2}b^+_2(D)P_+)\|\la'\>'$ $=0$ if and only if
$b^+\({d \over dx}\)\[e^x\D_0(x)+\D_1(x)-C\] = 0$ and
$W(z^{-2}b^+_2(D)P_+)\|\la\> =0$, respectivly.
This can be proved by using the facts that
$\mBRA W(z^m f(D)P_+),W(z^n g(D)P_+)\mKET$ $=0$ and
$\[W\(f(D)P_1+f(D+1)P_0)\),W(z^{-1}g(D)P_+)\] =0$.
Hence,
$$\eqalign{
W'(      b^+  (D-\la^1)P_-)\|\la'\>'=W(z     b^+(D+1)P_-)\|\la'\>' =0,\cr
W'(z^{-1}b^+_2(D-\la^1)P_+)\|\la'\>'=W(z^{-2}b^+_2(D)P_+)\|\la'\>' =0.
}$$
Therefore, the new characteristic polynomials are given by
$a'^-(w)=b^+(w-\la^1)$ and $b'^+(w)=b^+_2 (w-\la^1)$.

This completes the proof of Theorem 6.3.
\qed

%%%%%%%%%%%%%%%%%% The new char. poly.  %%%%%%%%%%%%%%%%%%%%%%%%%%%%%
\no {\bf 6.4.}~
The figure 2 illustrates the $W(P_\pm)$ part of the $\SWinf$ module:
(i), (ii) and (iii) correspond to the cases
$\la^1-\la^0=0$, $1$ and $-1$, respectively.

%%%%%%%%%%%%%%%%%%%% Fig. 2 %%%%%%%%%%%%%%%%%%%%%%%%%%%%%
\vskip10pt
\vbox{
\centerline{\epsfbox{swfig2.eps}}
\centerline{{\bf Figure 2.} Spectral Flow for the $\SWinf$ Algebra}
}%%%%%%%%%%%%%%%%%%  End of Fig. %%%%%%%%%%%%%
\no
The arrow $a$ corresponds to the generator $W(D^k P_-)$,
$b$ to $W(z^{-1} D^k P_+)$,
$c$ to \BR $W(z^{-1} D^k P_-)$,
$d$ to $W(z^{-2} D^k P_+)$,
$a'$ to $W'(D^k P_-)$, and
$b'$ to $W'(z^{-1} D^k P_+)$.
We can easily understand that
if $\la^1-\la^0=0$, then
$$
W'(      a'^-(w)P_-)=W(      a^-(w)P_-),\qquad
W'(z^{-1}b'^+(w)P_+)=W(z^{-1}b^+(w)P_+);
$$
if $\la^1-\la^0=1$, then
$$
W'(      a'^-(w)P_-)=W(z^{-1}b^-(w)P_-),\qquad
W'(z^{-1}b'^+(w)P_+)=W(      a^-(w)P_+);
$$
if $\la^1-\la^0=-1$, then
$$
W'(      a'^-(w)P_-)=W(z     b^+  (w+1)P_-),\qquad
W'(z^{-1}b'^+(w)P_+)=W(z^{-2}b^+_2(w  )P_+).
$$

%%%%%%%%%%%%%%%%%%%%%%%%%%%%%%%%%%%%%%%%%%%%%%
%%                                          %%
%%  7. Example: the $(B,C)$--system         %%
%%                                          %%
%%%%%%%%%%%%%%%%%%%%%%%%%%%%%%%%%%%%%%%%%%%%%%%%%%%%%%%%%%%%%%%%%%%%%%%
\bigbreak
\no{\bf 7. Example: the $(B,C)$--system}
\vskip2mm
%%%%%%%%%%%%%%%%%%%% BC system %%%%%%%%%%%%%%%%%%%%%%%%%%%%%%%%%%%%%%%%
\def\Eope{7.1}
\def\Evi {7.2}
\def\Evii{7.3}
\def\Efi {7.4}
\def\Efii{7.5}
\def\EBCbi{7.6}
\def\EBCbii{7.7}
\def\EBCdi{7.8}
\def\EBCdii{7.9}
%%%%%%%%%%%%%%%%%
\no{\bf 7.1.}~
In this section,
we give the free-field realization of $\SWinf$ %with central charge $C=1$
by using the $(B,C)$--system.
Here the superfields
$B(z,\t)=\b(z)+\t b(w)$ and $C(z,\t)= c(z)+\t\g(w)$
are defined by the following OPE:
$$
\g(z)\b(0) \sim -\b(z)\g(0) \sim {1\over z},\qquad
 c(z) b(0) \sim   b(z) c(0) \sim {1\over z},
\eqno(\Eope)$$
and the conformal weights of ($\b$, $\g$, $b$, $c$) are assigned as
($\la$+1, $-\la$, $\mu+1$, $-\mu$) with $\la,\mu\in\bC$.
Conformal dimension of $\t$ is thus $\la-\mu-1/2$.

%%%%%%%%%%%%%%%%%%%%% bosonization %%%%%%%%%%%%%%%%%%%%%%%%
\no{\bf 7.2.}~
For explicit calculation, it may be useful to
``bosonize'' the $(B,C)$--system as follows [FMS].
First we introduce free bosons $\p(x)$, $\sig(x)$ and
free fermions $\xi(z)$, $\eta(z)$ with the following OPE:
$$\eqalign{
\p(z)  \p(0) &\sim +\log z,\qquad
\sig(z)\sig(0)\sim -\log z,\cr
\eta(z)\xi(0)&\sim  \xi(z)\eta(0)\sim {1\over z}.\cr
}$$
Then $\b(z)$, $\g(z)$, $b(z)$ and $c(z)$ are expressed by
$\p(x)$, $\sig(x)$, $\xi(z)$ and $\eta(z)$ as
$$\eqalign{
\b(z)&\equiv :e^{-\sig(z)}:\partial\xi(z)=\sum_{n\in\bZ}\b_n z^{-n-\la-1},\cr
\g(z)&\equiv :e^{ \sig(z)}:       \eta(z)=\sum_{n\in\bZ}\g_n z^{-n+\la  },\cr
 b(z)&\equiv :e^{ \p  (z)}:              =\sum_{n\in\bZ} b_n z^{-n-\mu -1},\cr
 c(z)&\equiv :e^{-\p  (z)}:              =\sum_{n\in\bZ} c_n z^{-n+\mu   }.\cr
}$$
%Here $n$ runs over integers $(n\in\bZ)$.
It is easy to show that the OPE (\Eope) is
actually reproduced in this representation.

Let the mode expansions of $\sig(z)$ and $\p(z)$ be as follows:
$$\eqalign{
\sig(z)&=-\sum_{n\in\bZ_{\neq 0}}{\a_n\over n}z^{-n} +\a_0\log z +\sig_0,\cr
\p  (z)&=-\sum_{n\in\bZ_{\neq 0}}{ a_n\over n}z^{-n} + a_0\log z +\p_0.\cr
}$$
Introducing the bosonic vacuum $\|0\>$ satisfying
$\a_n\|0\>=a_n\|0\>=0$ for $n\ge 0$,
we define the $(\la,\mu)$--vacuum $\|\la,\mu\>$ by
$$
\|\la,\mu\>\equiv :e^{-\la\sig(0)-\mu\p(0)}:\|0\>=e^{-\la\sig_0-\mu\p_0}\|0\>.
\eqno(\Evi)$$
Note that it satisfies the following equation:
$$
\b_n\|\la,\mu\>=\g_{n+1}\|\la,\mu\>=
 b_n\|\la,\mu\>= c_{n+1}\|\la,\mu\>=0\qquad(n\ge 0).
\eqno(\Evii)$$

%%%%%%%%%%%%%%%%%%% characteristic polynomials %%%%%%%%%%%%%%%%%%%%%%%%
\no{\bf 7.3.}~
Here we discuss the free--field realization of
the fundamental representations of $\SWinf$, eqs.\ (\EDa) and (\EDc):
Recall that
eq.\ (\EDa) corresponds to the case where $a^-(w)=1$ and $b^+(w)=w-\la$,
while eq.\ (\EDc) to the case where $a^-(w)=w-\la$ and $b^+(w)=1$.

Using the correspondence in sect.\ 3.6,
we define the representation of $\SWinf$ over the $(B,C)$--system by
$$\eqalign{
W(z^n F(D))
\equiv&\oint{dz d\t \over 2\pi i} :B(z,\t) z^n f^A(D)P_A C(z,\t):\cr
=&\oint{dz\over 2\pi i}:\b(z) z^n f^0(D)\g(z):
 +\oint{dz\over 2\pi i} \b(z) z^n f^+(D) c(z) \cr
&+\oint{dz\over 2\pi i}  b(z) z^n f^-(D)\g(z)
 +\oint{dz\over 2\pi i}: b(z) z^n f^1(D) c(z):.\cr
}\eqno(\Efi)$$
Explicit calculation shows that the central charge $C=1$, and
$W(z^n F(D))$ has the following mode expansions:
$$\eqalign{
W(z^n F(D))
=&\sum_{\l\in\bZ} f^0(\la+\l):\b_{n+\l       }\g_{-\l}:
 +\sum_{\l\in\bZ} f^+(\mu +\l) \b_{n+\l-\la+\mu} c_{-\l} \cr
+&\sum_{\l\in\bZ} f^-(\la+\l)  b_{n+\l+\la-\mu}\g_{-\l}
 +\sum_{\l\in\bZ} f^1(\mu +\l): b_{n+\l       } c_{-\l}:.\cr
}\eqno(\Efii)$$
Thus, using eqs.\ (\Evii) and (\Efii), we can prove that
$\|\la,\mu\>$ is a highest weight vector if $\mu-\la=0$ or $1$:
$$\eqalign{
&W(z^n F(D)  )\|\la,\mu\>=0\qquad(n\ge1\,;\,{}^\forall F(w)\in\cA   ),\cr
&W(f  (D)P_+ )\|\la,\mu\>=0\qquad(          {}^\forall f(w)\in\bC[w]),\cr
&W(f^0(D)P_0+f^1(D)P_1)\|\la,\mu\>\in\bC\|\la,\mu\>\qquad
({}^\forall f^{0,1}(w)\in\bC[w]).
}$$

%%%%%%%%%%%%%%%%%%% characteristic polynomials %%%%%%%%%%%%%%%%%%%%%%%%
\no{\bf 7.4.}~
When $\mu-\la=0$, eq.\ (\Evii) implies that
$W(P_-)\|\la,\la\>=\sum_\l b_{-\l}\g_\l\|\la,\la\>=0$,
and thus we know that in this representation $a^-(w)=1$.
For characteristic polynomials $b^A_k(w)$ $(k\ge1\,;\,\AA)$,
one can prove the following equation as
in the free field realization of the $C=\pm1$ $\Winf$ algebra [M]:
$$
b^A_k(\la+\l)=0\qquad(\l=0,1,\cdots,k-1\,;\,\AA).
\eqno(\EBCbi)$$
As for $A=-$, for example,
eq.\ (\EBCbi) is obtained from the following equation
by setting $\l=0,1,\cdots,k-1$,
and picking up the coefficient of $b_{-(k-\l)}\|\la,\la\>$:
$$\eqalign{
0&=\[W\(z^{-k}b^-_k(D)P_-\),\b_{\l}\]\|\la,\la\>\cr
 &=b^-_k(\la+\l)\,b_{-(k-\l)}\|\la,\la\>.
}$$
Here we have used the fact that $b_{-n}\|\la,\la\>$ $(n\ge1)$ does not vanish.
Thus, solving eq.\ (\EBCbi),
we obtain the explicit form of characteristic polynomials:
$$
b^A_k(w)=(w-\la)(w-\la-1)\cdots(w-\la-k+1)\qquad (\AA).
\eqno(\EBCbii)$$
In particular, noticing that $b^+(w)~(=b^+_1(w))~=w-\la$,
we can rewrite eq.\ (\EBCbii) into the form
$b^A_k(w)=\prod_{\l=0}^{k-1}b^+(w-\l)$,
which is consistent with eq.\ (\Ecpk) for $a^-(w)=1$.

When $\mu-\la=1$,
we can similarly show that $a^-(w)=w-\la$ and $b^+(w)=1$.

%%%%%%%%%%%%%%%%%%% differential equations %%%%%%%%%%%%%%%%%%%%%%%%%%%%
\no{\bf 7.5.}~
The eigenvalue $\D_0(x)$ of the operator $-W\(e^{xD}P_0\)$ is
calculated as follows [M]:
$$\eqalign{
-W&\(e^{xD}P_0\)\|\la,\mu\>\cr
&=-\oint{dz\over 2\pi i}:\b(z)e^{xD}\g(   z):\|\la,\mu\>\cr
&=-\oint{dz\over 2\pi i}:\b(z)      \g(e^xz):\|\la,\mu\>\cr
&=-\oint{dz\over 2\pi i}
\[:e^{-\sig(z)}:\partial\xi(z):e^{\sig(e^xz)}:\eta(e^xz)+{1\over z-e^xz}\]
:e^{-\la\sig(0)-\mu\p(0)}:\|0\>\cr
&={-1\over e^x-1}\oint{dz\over 2\pi i}{1\over z}
\[:e^{-\sig(z)+\sig(e^xz)}:-1\]:e^{-\la\sig(0)-\mu\p(0)}:\|0\>\cr
&=-\,{e^{\la x}-1\over e^x-1}\|\la,\mu\>.
}$$
Namely, we obtain
$$
\D_0(x)=-\,{e^{\la x}-1\over e^x-1}.
\eqno(\EBCdi)$$
Similarly we can calculate $\D_1(x)$ as
$$\eqalign{
-W\(e^{xD}P_1\)\|\la,\mu\>
&=-\oint{dz\over 2\pi i}:b(z)c(e^xz):\|\la,\mu\>\cr
&=+\,{e^{\mu x}-1\over e^x-1}\|\la,\mu\>,
}$$
{\it i.e.},
$$
\D_1(x)=+\,{e^{\mu x}-1\over e^x-1}.
\eqno(\EBCdii)$$

%%%%%%%%%%%%%%%%%%%% characteristic polynomials %%%%%%%%%%%%%%%%%%%%%%%%
\no{\bf 7.6.}~
If $\mu-\la=0$, then the generating functions $\D_a(x)$ $(a=0,1)$ given in
eqs.\ (\EBCdi) and (\EBCdii)
actually satisfy the differential equations in the previous section
with $a^-(w)=1$, $b^+(w)=w-\la$ and $C=1$:
$$\eqalign{
&~~~~\D_0(x)+\D_1(x)=0,\cr
\({d\over dx}-\la\) &\[e^x\D_0(x) + \D_1(x)-1\]=0.
}$$
If $\mu-\la=1$, then they satisfy the differential equations
with $a^-(w)=w-\la$, $b^+(w)=1$ and $C=1$.

%%%%%%%%%%%%%%%%%%%%%%%%%%%%%%%%%%%%%%%%%%%%%%
%%                                          %%
%%  8. Conclusion and Discussion            %%
%%                                          %%
%%%%%%%%%%%%%%%%%%%%%%%%%%%%%%%%%%%%%%%%%%%%%%%%%%%%%%%%%%%%%%%%%%%%%%%%%%%
\bigbreak
\no{\bf 8. Conclusion and Discussion}
\vskip2mm
%%%%%%%%%%%%%%%%%%%%%%%%%%%%%%%%%%%%%%%%%%%%%%%%%%%%%%%%%%%%%%%%%%%%%%%%%%%
\no
In this paper we have formulated the super $\Winf$ algebra, $\SWinf$,
as a central extension of the Lie super
algebra of superdifferential operators acting on the polynomial
algebra over $2 \times 2$ supermatrices.
We then have studied the quasifinite highest weight modules over
$\SWinf$. Our discussion is parallel with Kac and Radul's one.
The quasifiniteness of the modules is characterized by polynomials,
and the generating functions of highest weights, $\D_a(x)$ $(a=0,1)$,
satisfy a set of differential equations.

Mathematically, there are many things to be clarified.
In the bosonic counterpart, we have already
obtained the determinant formulae of the
$\Winf$ module and the character formulae of
the degenerate representations [AFOQ,AFMO1,AFMO2].
Furthermore, we study the structure of subalgebras of
the $\Winf$ algebra [AFMO3], especially, the $\Win$ algebra
(algebra without spin one current).
The supersymmetric extension of these analysis seems to be of some interest.

Since $\SWinf$ contains the $N=2$ superconformal algebra as a
subalgebra, $\SWinf$ has another interesting application, {\it geometry}.
In fact, geometry of complex manifolds
(especially the Calabi--Yau manifolds and their mirrors),
and topological field theory have been studied
by using the $N=2$ superconformal algebra.
For example,
the Calabi--Yau manifolds are described by the $N=2$ supersymmetric
non-linear $\sigma$ model or by Landau--Ginzburg orbifolds,
whose elliptic genera have been computed recently in refs.\ [EOTY, KYY].
$\SWinf$ naturally appears there through the free field realization,
and thus, using $\SWinf$ we may obtain more information
than using the $N=2$ superconformal algebra only.
We hope to report on these subjects in our future communication.

Finally we comment on the family of (super) $\cW$ infinity algebras.
The super $\Win$ algebra given in ref.\ [BPRSS] is a subalgebra of $\SWinf$,
which corresponds to the relation between $\Winf$ and $\Win$ [AFMO3].
The super $\Win$ $(\Win^{1,1})$ was extended to $\Win^{M,N}$ [O].
Similarly, by replacing $2 \times 2$ supermatrices by
$(M+N) \times (M+N)$ supermatrices,
one can easy extend $\SWinf$ $(\Winf^{1,1})$ to $\Winf^{M,N}$,
which contains $\Win^{M,N}$ as a subalgebra.

%%%%%%%%%%%%%%%%%%%%%%%%%%%%%%%%%%%%%%%%%%%%%%
%%                                          %%
%%  Acknowledgements                         %%
%%                                          %%
%%%%%%%%%%%%%%%%%%%%%%%%%%%%%%%%%%%%%%%%%%%%%%%%%%%%%%%%%%%%%%%%%%%%%
\bigbreak
\no{\bf Acknowledgements}
\vskip2mm
%%%%%%%%%%%%%%%%%%%%%%%%%%%%%%%%%%%%%%%%%%%%%%%%%%%%%%%%%%%%%%%%%%%%%%%%%
\no
Two of the authors (Y.M. and S.O.) would like
to thank members of YITP for their hospitality.
This work is supported in part by
Grant-in-Aid for Scientific Research from Ministry of
Science and Culture, and by Soryushi-Shogakkai.

%The authors would like to thank ??? for valuable discussions.
%H.A. is partly supported by Soryushi-syougakkai.

%%%%%%%%%%%%%%%%%%%%%%%%%%%%%%%%%%%%%%%%%%%%%%
%%                                          %%
%%  A. Appendix                             %%
%%                                          %%
%%%%%%%%%%%%%%%%%%%%%%%%%%%%%%%%%%%%%%%%%%%%%%%%%%%%%%%%%%%%%%%%%%%%%%
\bigbreak
\no{\bf Appendix A.~ Embedding into $\Glinf$}
\vskip2mm
%%%%%%%%%%%%%%%%%%%%%%%%% gl inf %%%%%%%%%%%%%%%%%%%%%%%%%%%%%%%%%%%%%%
\def\Eem{A.1}
%%%%%%%%%%%%%%%
\no{\bf A.1.}~
Let $E_{mn}$ $(m,n\in\bZ)$ denote the matrix unit of infinite size:
$E_{mn}=(\d_{im}\d_{jn})_{i,j\in\bZ}$.
An infinite dimensional Lie algebra $\gl$ is then defined as
$$
\gl\equiv\set{\sum_{m,n\in\bZ}a_{mn}E_{mn}}{
a_{mn}=0\quad{\rm for}\quad |m-n|\gg 0}.
$$
We further define $\glinf\equiv\gl\otimes_\bC \cA$
with $\cA$ the algebra over $(2\times 2)$ supermatrices
$$
\cA\equiv\set{\Fgmat{f}{m}}{f^A(m)\in\bC[m]\,;\,\AA },
$$
with the same $\bZ_2$--gradation as eq.\ (\EA).
We have changed the arrangement of matrix elements from eq.\ (\EA)
for later convenience.

%%%%%%%%%%%%%%%%%%%%%%%%% embeding %%%%%%%%%%%%%%%%%%%%%%%%%%%%%%%
\no{\bf A.2.}~
Let $\t$ be a Grassmann number and $z^n F(D)=\sum_A z^n f^A(D)P_A\in\swinf$.
The embedding map $\vp:\swinf\h->\glinf$ is defined through the action
of $\swinf$ on $\bC[z,z^{-1}]\oplus\bC[z,z^{-1}]\t$ as follows:
$$
z^n F(D)\cdot z^m (1,\t)\equiv
\sum_{\l\in\bZ} z^\l (1,\t)\cdot\vp\(z^n F(D)\)_{\l,m}.
\eqno(\Eem)$$
Since the action of $P_A$'s on $(1,\t)$ is given as
$$
\Smat{P_1 & P_- \cr P_+ & P_0 \cr}\cdot 1 =\Smat{1 & 0 \cr \t & 0 \cr},\qquad
\Smat{P_1 & P_- \cr P_+ & P_0 \cr}\cdot \t=\Smat{0 & 1 \cr  0 & \t\cr},
$$
the left hand side of eq.\ (\Eem) reduces to
$$
z^{m+n}(1,\t)\cdot\Fgmat{f}{m}
=\sum_\l z^\l (1,\t)\cdot\d_{\l,m+n}\Fgmat{f}{m}.
$$
Thus, we obtain
$$
\vp\(z^n F(D)\)_{\l,m}
=\d_{\l,m+n}\Fgmat{f}{m}
\equiv\(\La^n[F(d)]\)_{\l,m}.
$$
Here $\La$ and $[F(d)]$ stand for the following infinite matrices:
$$
\La=\sum_{m\in\bZ}E_{m,m-1},\qquad %\La=(\d_{m-1,n})_{m,n\in\bZ}
[F(d)]=\sum_{m\in\bZ}F(m)E_{mm},\qquad %[F(d)]=(F(m)\d_{mn})_{m,n\in\bZ}
F(m)=\Fgmat{f}{m}.
$$
By definition, the map $\vp:\swinf\->\glinf$ is homomorphic,
{\it i.e.}, $$\vp(AB)=\vp(A)\vp(B).$$

%%%%%%%%%%%%%%%%%%%%%%%%% principal gl inf %%%%%%%%%%%%%%%%%%%%%%%%
\no{\bf A.3.}~
Let us introduce new variables
$\mu=\t+z\partial_\t$ and $\mu^{-1}=z^{-1}\t+\partial_\t$.
Note that they satisfy $\mu^2=z$, $\mu\mu^{-1}=\mu^{-1}\mu=1$,
and also that
$$
\Smat{P_1 &      P_- \cr    P_+ & P_0 \cr}=
\Smat{\hfill P_1 & \mu^{-1}P_0 \cr \mu P_1 & \hfill P_0 \cr}.
$$
Thus, we can think of
the diagonal elements $P_0$ and $P_1$ as the fundamental elements.
Hence, elements of $\glinf$ can be represented by matrices
with half--integer indices as follows:
$$
\glinf=\set{\sum_{\a,\b\in\bZ/2}a_{\a\b}E_{\a\b}}{
a_{\a\b}=0\quad{\rm for}\quad |\a-\b|\gg 0 },
$$
where we denote $E_{\a\b}=(\d_{\mu\a}\d_{\nu\b})_{\mu,\nu\in\bZ/2}$.
The $\bZ_2$--gradation is assigned as
$\glinf=\glinf^{(0)}\oplus\glinf^{(1)}$
and $E_{\a\b}\in\glinf^{(0)}$ if and only if $\a-\b\in\bZ$,
otherwise $E_{\a\b}\in\glinf^{(1)}$.

%%%%%%%%%%%%%%%%%%%%% affine gl inf %%%%%%%%%%%%%%%%%%%%%%%%%%%
\no{\bf A.4.}~
Denoting by $\tW(A)$ the element in $\Glinf$
which corresponds to an element $A$ in $\glinf$,
we introduce $\Glinf$ as the central extension of $\glinf$
with the following (anti--) commutation relation:
$$
\[\tW(A),\tW(B)\mKET\equiv \tW(\[A,B\mKET)- C\tPsi(A,B),\qquad
\tPsi(A,B)=\str J\[A,B\mKET,
$$
where $J=\sum_{\a\ge 0}E_{\a\a}$ and
$\str A=-\sum_{\a\in\bZ/2}(-1)^{2\a}(A)_{\a\a}$.
The fundamental (anti--) commutation relation for $\Glinf$ is
$$\eqalign{
\[\tW(E_{\a\b}),\tW(E_{\g\d})\mKET
&=\d_{\b\g}\tW(E_{\a\d})-(-1)^{2(\a-\b)2(\g-\d)}\d_{\d\a}\tW(E_{\g\b})\cr
&~\hskip.5truein+C\d_{\b\g}\d_{\d\a}(-1)^{2\a}\(\t(\a)-\t(\g)\),
%\sum_{2\mu\in\bZ_{\ge0}}\(\d_{\a\mu}-\d_{\b\mu}\).
}$$
where $\t(\a)=1$ if $\a\ge 0$ and otherwise $\t(\a)=0$.
The (anti--) commutation relation for $\SWinf$ embedded in $\Glinf$
is thus given by
$$\eqalign{
&\[\tW(\vp\(z^n F(D)\)),\tW(\vp\(z^m G(D)\))\mKET\equiv
\tW\(\[\vp\(z^n F(D)\),\vp\(z^m G(D)\)\mKET\)\cr
&~\hskip1truein+C\d_{n+m,0}\(\tsum_{j\ge0}-\tsum_{j\ge n}\)
\left\{ h^{11}(j)+h^{-+}(j)-h^{+-}(j)-h^{00}(j)\right\},\cr
}$$
with $h^{A,B}(j)=f^A(j+m)g^B(j)$,
which is the same as that in eq.\ (\Ecri).

%%%%%%%%%%%%%%%%%%%%%%%%% spectral flow %%%%%%%%%%%%%%%%%%%%%%%%%
\no{\bf A.5.}~
We can easily understand automorphisms of $\SWinf$
in eq.\ (\Esf) as the ones of $\Glinf$ as follows.

Since any automorphisms $\pi_{w}:\swinf$ $\->\swinf$ and
$\pi_{g}:\glinf\->\glinf$ are realized by the basis transformation
$\pi_{z}:\bC[z,z^{-1},\t]\->\bC[z,z^{-1},\t]$ as
$$\eqalign{
z^n F(D)\cdot \pi_z\(z^m (1,\t)\)&=
\sum_\l \pi_z\(z^\l (1,\t)\)\cdot\pi_{g}\(\vp\(z^n F(D)\)\)_{\l,m},\cr
\pi_{w}\(z^n F(D)\)\cdot \pi_z^{-1}\(z^m (1,\t)\)&=
\sum_\l \pi_z^{-1}\(z^\l (1,\t)\)\cdot\vp\(z^n F(D)\)_{\l,m},
}$$
we have
$$
\pi_{w}\(z^n F(D)\)\cdot z^m (1,\t)=
\sum_\l z^\l (1,\t)\cdot\pi_{g}\(\vp\(z^n F(D)\)\)_{\l,m}.
$$

Furthermore, one can also easily obtain the induced automorphisms
$\pi_{w}\!:\SWinf\!\-> \SWinf$ and $\pi_{g}:\Glinf\->\Glinf$
with some modifications coming from central terms.
For example, for the transformation
$\pi_z\(z^m (1,\t)\)=z^m (z^{\la^1},z^{\la^0}\t)$,
the automorphisms are given by
$$\eqalign{
&\pi_{g}\(\tW\(\vp\(z^n F(D)\)\)\)\cr
&~\hskip.4truein=\tW\(
\Smat{\d_{\l,m+n}            f^1(m+\la^1)
     &\d_{\l,m+n+\la^0-\la^1}f^-(m+\la^0)\cr
      \d_{\l,m+n+\la^1-\la^0}f^+(m+\la^1)
     &\d_{\l,m+n}            f^0(m+\la^0)\cr}_{\l,m\in\bZ}\)\cr
&~\hskip1truein +C\d_{n,0}
\mBRA\(\tsum_{j\ge\la^1}-\tsum_{j\ge0}\)f^1(j)
   - \(\tsum_{j\ge\la^0}-\tsum_{j\ge0}\)f^0(j)\mKET,\cr%&\cr
}$$
$$\eqalign{\pi_{w}\(W\(z^n F(D)\)\)
&=W\(z^n\{f^1(D+\la^1)P_1 + z^{\la^0-\la^1}f^-(D+\la^0)P_-\right.\cr
&~~~~~~~\left.+ z^{\la^1-\la^0}f^+(D+\la^1)P_+ + f^0(D+\la^0)P_0\}\)\cr
&~+C\d_{n,0}
\mBRA\(\tsum_{j\ge\la^1}-\tsum_{j\ge0}\)f^1(j)
   - \(\tsum_{j\ge\la^0}-\tsum_{j\ge0}\)f^0(j)\mKET,\cr
}$$
which corresponds to the spectral flow in eq.\ (\Esf).

%%%%%%%%%%%%%%%%%%%%% quasifinite %%%%%%%%%%%%%%%%%%%%%%%%%%%%%%%%
\no{\bf A.6.}~
We can reformulate the quasifinite highest weight representation
of $\SWinf$ in terms of $\Glinf$.
We denote $\tA_\a=M^{-2\a}A_\a$ with a diagonal matrix $A_\a$
and $M=\sum_{\a\in\bZ/2}E_{\a,\a-1/2}$, $\La=M^2$.

We define
$I_\a\equiv \set{\tA_\a}{\tW(\tA_\a)\widetilde{\|\la\>}=0}$,
where $\widetilde{\|\la\>}$ is the highest weight vector such that
$\tW(\tA_\a)\widetilde{\|\la\>}=0$ for all $\tA_\a$ with $\a>0$.
%and $\tW(\tA_0)\widetilde{\|\la\>}=\tA_0\widetilde{\|\la\>}$.
Then we can show that $I_\a$ is an ideal, {\it i.e.},
if $\tA_\a\in I_\a$, then $H \tA_\a\in I_\a$ for any diagonal matrix $H$.
Hence, $I_\a$ is generated by a characteristic matrix $\tC_\a$, {\it i.e.},
if $\tA_\a\in I_\a$,
then there exists a diagonal matrix $H$ such that $\tA_\a=H \tC_\a$.
The relation between $C_\a$ and
the characteristic polynomials $a^-(w)$ and $b^A_k(w)$ in eq.\ (\Ecp) is
$$
\Smat{(C_{n    })_{kk} & (C_{n+\ha})_{k+\ha,k+\ha} \cr
      (C_{n-\ha})_{kk} & (C_{n    })_{k+\ha,k+\ha} \cr}=
\Smat{b^1_n(k) & b^-_n(k) \cr
      b^+_n(k) & b^0_n(k) \cr},
$$
where $n\ge 0$, $k\in\bZ$,
and we set $b^-_0(k)=a^-(k)$, $b^0_0(k)=b^1_0(k)=b^+_0(k)=0$.
The matrix $\(\sum_{\a>0}\tC_{\a}\)_{\mu,\nu\in\bZ/2}$
is arranged explicitly as follows:
%%%%%%%%%%%%%%%%% def %%%%%%%%%%%%%%%%%%%%
\def\UB#1{\mathop{{}\atop{#1}}_\smile}
\def\M{\matrix}
\def\MB#1#2{\M{b^1_{#2}(#1)&b^-_{#2}(#1)\cr b^+_{#2}(#1)&b^0_{#2}(#1)\cr}}
\def\MA#1  {\M{~~~~0~~~~   &a^-     (#1)\cr ~~~~0~~~~   &~~~~0~~~~   \cr}}
\def\MAZ   {\M{~~~~0~~~    &a^-     ( 0)\cr ~~~~0~~~    &~~~~0~~~    \cr}}
\def\MZ    {\M{~~~~0~~~~   &~~~~0~~~~   \cr ~~~~0~~~~   &~~~~0~~~~   \cr}}
\def\DD{\ddots}
%%%%%%%%%%%%%%%%%%%%%%%%%%%%%%%%%%%
$$\eqalign{
&{~~\mu\backslash\nu~~~\cdots
{}~~~~~~       \UB{-2}~~~~   \UB{-{3\over2}}
{}~~~~~~~~~~~  \UB{-1}~~~~   \UB{-\ha}
{}~~~~~~~~~~~~~\UB{0 }~~~~~~~\UB{ \ha}~~~~~\cdots}\cr
&\M{\vdots\cr -3)\cr -{5\over2})\cr\cr
              -2)\cr -{3\over2})\cr\cr
              -1)\cr -      \ha)\cr\cr
               0)\cr        \ha)\cr\vdots\cr}
\left[\M{
\M{\DD\cr} &\M{ & \cr} &\M{ & \cr} &\M{ & \cr} &\M{   \cr}\cr
\M{\cr\cr} &\MB{-2}{ } &\MB{-1}{2} &\MB{ 0}{3} &\M{\cr\cr}\cr
\strut     &           &           &           &          \cr
\M{\cr\cr} &\MA{-2}    &\MB{-1}{ } &\MB{ 0}{2} &\M{\cr\cr}\cr
\strut     &           &           &           &          \cr
\M{\cr\cr} &\MZ        &\MA{-1}    &\MB{ 0}{ } &\M{\cr\cr}\cr
\strut     &           &           &           &          \cr
\M{\cr\cr} &\MZ        &\MZ        &\MAZ       &\M{\cr\cr}\cr
\M{   \cr} &\M{ & \cr} &\M{ & \cr} &\M{ & \cr} &\M{\DD\cr}\cr
}\right].
}$$

We can show that there exist diagonal matrices $H_n$ $(n=1,2,3)$ such that
$$\eqalign{
\tC_{\a+\ha}&=(M^{-1} H_1) \tC_\a = \tC_\a(M^{-1} H_2),\cr
\tC_\a \tC_\b &= H_3 \tC_{\a+\b}.
}$$
This equation can be solved recursively, and we obtain
$$\eqalign{
&\l.c.m.\((\tC_{\ha})_{\mu,\mu+\ha},(\tC_{\ha})_{\mu+\ha,\mu+1},\cdots,
(\tC_{\ha})_{\mu+\a-\ha,\mu+\a}\)
{}~~~\vert~~~(\tC_{\a})_{\mu,\mu+\a},\cr
&(\tC_{\a})_{\mu,\mu+\a}~~~\vert~~~
(\tC_{\ha})_{\mu,\mu+\ha}(\tC_{\ha})_{\mu+\ha,\mu+1}\cdots
(\tC_{\ha})_{\mu+\a-\ha,\mu+\a}.
}$$
for all $\mu$, $\a\in\bZ/2$.
This is equivalent to the relations in eqs.\ (\Eai)--(\Ebii).
Furthermore, if the elements of $\tC_{\ha}$ are mutually prime,
then we have the relation $\tC_\a=(\tC_{\ha})^{2\a}$,
which is the same as eq.\ (\Ecpk).
Note that if we set
$$
(\tC^{\l cm}_\a)_{\mu,\mu+\a}
=\l.c.m.\((\tC_{\ha})_{\mu,\mu+\ha},(\tC_{\ha})_{\mu+\ha,\mu+1},\cdots,
(\tC_{\ha})_{\mu+\a-\ha,\mu+\a}\),
$$
then one may show that
$\tW(\tC^{\l cm}_\a)\widetilde{\|\la\>}$ is a null state.
We will prove it in Appendix B.

%%%%%%%%%%%%%%%%%%%%%%%%%%%%%%%%%%%%%%%%%%%%%%%%%%%%%%%%%%%%%%%%%%%%%%%
\bigbreak
\no{\bf Appendix B.~ Null Vector Condition}
\vskip2mm
%%%%%%%%%%%%%%%%%%%% null vector %%%%%%%%%%%%%%%%%%%%%%%%%%%%%%%%%%%%%%%
\def\FChi{
Note that the whole algebra $\SWinf$ is generated by
$W(P_\pm)$, $W(z^{\pm 1}P_\mp)$ and $W(DP_0)$.}
\def\FChii{
In the bosonic case,
the $\Winf$ is generated by $W(z^{\pm 1})$ and $W(D^2)$.
To show the null vector condition,
it is sufficient that $W(ze^{x(D+1)})\|\chi\>$
is a null vector or vanishes.}
\def\Elcm{B.1}
%%%%%%%%%%%%%%%%%
\no{\bf B.1.}~
We discussed the quasifinite highest weight module
as the {\it generalized Verma module} [KR],
which is annihilated by the parabolic subalgebra.
However, as seen in Corollary 4.5,
the characteristic polynomials $b^A_k(w)$ are not fixed uniquely.
Here we will show that
the characteristic polynomials are uniquely determined
if we demand that the quasifinite highest weight module be {\it irreducible}.

We first introduce a bilinear form.
Recall that $V(\la)$ is the Verma module over $\SWinf$,
generated by the highest weight vector $\|\la\>$, such that
$$
W(        D^k P_+)\|\la\>=0,\qquad
W(z^{n+1} D^k P_A)\|\la\>=0,\qquad
W(      e^{xD}P_a)\|\la\>=-\D_a(x)\|\la\>
$$
with $n,k\in\bZ_{\ge 0}$, $A=0,1,\pm$ and $a=0,1$.
The dual module $V(\la)^*$ is generated by $\<\la\|$ which satisfies
$$\<\la\| W(         D^k P_-)=0,\qquad
\<\la\| W(z^{-n-1} D^k P_A)=0,\qquad
\<\la\| W(       e^{xD}P_a)=-\D_a(x)\<\la\|
$$
with $n,k\in\bZ_{\ge 0}$, $A=0,1,\pm$ and $a=0,1$.
The bilinear form
$V(\la)^* \otimes V(\la)\->\bC$ is uniquely defined by
$\<\la\|\la\> = 1$ and
$\(\< u\| W\)\| v\>=\< u\|\( W\| v\>\)$ for any
$\< u\|\in V(\la)^*$, $\| v\>\in V(\la)$ and $W\in\SWinf$.

The null vector $\|\chi\>$ is defined by the condition that
$\< u\|\chi\>=0$ for all $\< u\|\in V(\la)^*$.

%%%%%%%%%%%%%%%%%%% LCM Theorem %%%%%%%%%%%%%%%%%%%%%%%%%%%%%%%%%%%%%%
\no{\bf B.2.}~
We let $b^-_0(w)=a^-(w)$, $b^A_0(w)=0$ with $A=0,1,+$ and
$$\eqalign{
b^+_k(w)&=\l.c.m.\(b^+(w),a^-(w-1),b^+(w-1),a^-(w-2),\cdots,b^+(w-k+1)\),\cr
b^0_k(w)&=\l.c.m.\(a^-(w),b^+(w  ),a^-(w-1),b^+(w-1),\cdots,b^+(w-k+1)\),\cr
b^1_k(w)&=\l.c.m.\(b^+(w),a^-(w-1),b^+(w-1),a^-(w-2),\cdots,a^-(w-k  )\),\cr
b^-_k(w)&=\l.c.m.\(a^-(w),b^+(w  ),a^-(w-1),b^+(w-1),\cdots,a^-(w-k  )\),\cr
}\eqno(\Elcm)$$
for $k\in\bZ_{>0}$.
We will show the following Theorem:

%%%%%%%%%%%%%%%%% Theorem %%%%%%%%%%%%%%%%%%%%%%%%%%%%%%%
\no{\bf\csc Theorem.}~{\it
If the weight functions $\D_0(x)$ and $\D_1(x)$ satisfy
the differential equation {\rm (\Ediv)},
then $\|\chi^A_k\>\equiv W(z^{-k}e^{yD}b^A_k(D) P_A)\|\la\>$
is a null vector for all $y\in\bC$, $k\in\bZ_{\ge0}$ and $A\in\{0,1,+,-\}$.
}%%%%%%%%%% end of the theorem %%%%%%%

\no
To obtain the quasifinite {\it irreducible} highest weight module,
we must factor out the null vectors
which are characterized by the polynomials in eq.\ (\Elcm).
Since
there possibly exist additional null vectors
for some special values of $C$ and $\D_a(x)$,
we here discuss the generic case.

%%%%%%%%%%%%%%%%% Proof of the Theorem %%%%%%%%%%%%%%%%%%%%%%%%%%%%%%%%%%%
\no{\bf B.3.}~
{\it Proof of the Theorem.}

\no
First we have the following Lemma:

%%%%%%%%%%%%%%%%%%%% Lemma %%%%%%%%%%%%%%%%%%%%%%%%
\no{\bf\csc Lemma.}~{\it
The subalgebra
$$
\SWinf^+\equiv
\mBRA W(D^k P_+),W(z^{n+1} D^k P_A)\,\|\,{n,k\in\bZ_{\ge0},\,A=0,1,\pm}\mKET
$$
of $\SWinf$ is generated by $W(D^k P_+)$ and $W(z D^k P_-)$
with $k\in\bZ_{\ge0}$.}{\foot{$^\dagger$}\FChi}
%%%%%%%%%% end of the Lemma %%%%%%%%%%%%

\no{\it Proof }.
$W(z D^k P_a)$ with $a=0,1$ are obtained as follows:
$$\eqalign{
W(z D^k P_0)&=\{W(D P_+),W(z D^k P_-)\}-\{W(P_+),W(z D^{k+1} P_-)\},\cr
W(z D^k P_1)&=\{W(P_+),W(z D^k (D+1)P_-)\}-\{W(D P_+),W(z D^k P_-)\}.
}$$
One can further obtain $W(z^{n}D^k P_A)$ from $W(z^{n-1}D^\l P_A)$
by taking (anti--) commutators with $W(z P_0)$ or $W(z P_1)$.
\qed

\no
Hence, to prove that $\|\chi\>$ is a null vector,
it is sufficient to show that
$W(e^{xD}P_+)\|\chi\>$ and $W(z e^{x(D+1)}P_-)\|\chi\>$
are null vectors or vanish for all $x\in\bC$.{\foot{$^\ddagger$}\FChii}
The proof of the Theorem is given by induction as follows:

%%%%%%%%%%%%%%% induction I %%%%%%%%%%%%%%%%%%%%%%%%%%%%%%%%
\no{\bf\csc Step 1.}~{\it
$\|\chi^-_0\>$ and $\|\chi^+_1\>$ are null vectors.
}%%%%%%%%%% end of the Lemma %%%%%%%%%%%%

\no{\it Proof }.
{}From the differential equation for $\D_0(x)$ and $\D_1(x)$, we obtain
$$\eqalign{
W(e^{xD}P_+)\|\chi^-_0\>
&=a^-\({d\over dX}\) \[   \D_0(X) + \D_1(X) \]\|\la\>=0,\cr
W(z e^{x(D+1)}P_-)\|\chi^+_1\>&=
b^+\({d\over dX}\) \[\D_1(X) + e^X\D_0(X) - C \]\|\la\>=0,\cr
}$$
with $X\equiv x+y$.
Moreover, $W(z e^{x(D+1)}P_-)\|\chi^-_0\>=0$ and
$W(e^{xD}P_+)\|\chi^+_1\>=0$.
\qed

%%%%%%%%%%%%%%% induction II %%%%%%%%%%%%%%%%%%%%%%%%%%%%%%
\no{\bf\csc Step 2.}~{\it
If $\|\chi^-_{k-1}\>$ and $\|\chi^+_k\>$
are null vectors for a positive integer $k$, then
$\|\chi^0_k\>$ and $\|\chi^1_k\>$ are also null vectors.
}%%%%%%%%%% end of the Lemma %%%%%%%%%%%%

\no{\it Proof }.
Since $b^+_k(w)\,\|\,b^0_k(w),\,b^1_k(w)$,
and $b^-_{k-1}(w)\,\|\,b^0_k(w),\,b^1_k(w+1)$,
the following four vectors are null (here $X\equiv x+y$):
$$\eqalign{
W(e^{xD}P_+)\|\chi^0_k\>&=-W(z^{-k}e^{XD}b^0_k(D)P_+)\|\la\>,\cr
W(e^{xD}P_+)\|\chi^1_k\>&= e^{-kx} W(z^{-k}e^{XD}b^1_k(D)P_+)\|\la\>,\cr
W(z e^{x(D+1)}P_-)\|\chi^0_k\>&=
e^{(-k+1)x} W(z^{-k+1}e^{XD}b^0_k(D)P_-)\|\la\>,\cr
{}~\hskip.9truein
W(z e^{x(D+1)}P_-)\|\chi^1_k\>&=
-W(z^{-k+1}e^{X(D+1)}b^1_k(D+1)P_-)\|\la\>.
{}~\hskip.3truein\qed
}$$

%%%%%%%%%%%%%%% induction III %%%%%%%%%%%%%%%%%%%%%%%%%%%%%
\no{\bf\csc Step 3.}~{\it
If $\|\chi^0_k\>$ and $\|\chi^1_k\>$ are null vectors
for a positive integer $k$, then
$\|\chi^-_k\>$ and $\|\chi^+_{k+1}\>$ are also null vectors.
}%%%%%%%%%% end of the Lemma %%%%%%%%%%%%

\no{\it Proof }.
Since $b^0_k(w),\,b^1_k(w)\,\|\,b^-_{k}(w)$ and
$b^1_{k}(w)\,\|\,b^+_{k+1}(w)$ and $b^0_{k}(w)\,\|\,b^+_{k+1}(w+1)$,
the following two vectors are null:
$$\eqalign{
W(e^{xD}P_+)\|\chi^-_k\>&=
W(z^{-k}(e^{-kx}e^{XD}b^-_k    (D)P_0+e^{XD}    b^-_k    (D)  P_1))\|\la\>,\cr
W(z e^{x(D+1)}P_-)\|\chi^+_{k+1}\>&=
W(z^{-k}(e^{-kx}e^{XD}b^+_{k+1}(D)P_1+e^{X(D+1)}b^+_{k+1}(D+1)P_0))\|\la\>.\cr
}$$
Moreover, $W(e^{xD}P_+)\|\chi^+_k\>=0$ and
$W(ze^{x(D+1)}P_-)\|\chi^-_{k+1}\>=0$.
\qed

Thus we have completed the proof of Theorem B.1.%\qed

%\vfill\eject
%%%%%%%%%%%%%%%%%%%%%%%%%%%%%%%%%%%%%%%%%%%%%%
%%                                          %%
%%  References                              %%
%%                                          %%
%%%%%%%%%%%%%%%%%%%%%%%%%%%%%%%%%%%%%%%%%%%%%%%%%%%%%%%%%%%%%%%%%%%%%%%%
\vskip 3mm
\no{\bf References}
\vskip2mm
%%%%%%%%%%%%%%%%%%%%%%% reference %%%%%%%%%%%%%%%%%%%%%%%%%%%%%%%%%%
\item{ }{ }
\vskip-0.5truecm
\baselineskip=14pt

\itemitem{[AFMO1]}{Awata, H., Fukuma, M., Matsuo, Y., Odake, S.:
{\it Determinant Formulae of Quasi-Finite Representation
of $\Winf$ Algebra at Lower Levels},
Preprint \BR
YITP/K-1054, UT-669, SULDP-1994-1, January 1994,(hep-th/9402001)}

\itemitem{[AFMO2]}{Awata, H., Fukuma, M., Matsuo, Y., Odake, S.:
 {\it Determinant and Full Character Formulae of
 Quasi-Finite Representation of $\Winf$ Algebra},
 %in preparation}
 Preprint YITP/K-1060, UT-672, SULDP-1994-3, to appear}%March 1994}

\itemitem{[AFMO3]}{Awata, H., Fukuma, M., Matsuo, Y., Odake, S.:
in preparation}
%Preprint YITP/K-10??, UT-6??, SULDP-1994-4, February 1994}

\itemitem{[AFOQ]}{Awata, H., Fukuma, M., Odake, S., Quano, Y.-H.:
{\it Eigensystem and Full Character Formula of the $\Winf$ Algebra
with $c=1$},
Preprint YITP/K-1049, SULDP-1993-1, RIMS-959, December 1993,
(hepth/9312208)}, to be published in Lett. Math. Phys.

\itemitem{[B]}{Bakas, I.:
{\it The large--N limit of Extended Conformal Systems},
Phys. Lett. {\bf 228B}, 57-63 (1989)}

\itemitem{[BdWV]}{Bergshoeff, E., de Wit, B., Vasiliev, X.:
{\it The Structure of the Super--$W_{\infty}(\lambda)$ Algebra},
Nucl. Phys. {\bf B366}, 315-346 (1991)}

%\itemitem{[BK1]}{I. Bakas and E. Kiritsis,
%{\it BOSONIC REALIZATION OF A UNIVERSAL W ALGEBRA AND
%Z(INFINITY) PARAFERMIONS.}
%Nucl. Phys. {\bf B343}, 185-204 (1990);
%{GRASSMANNIAN COSET MODELS AND UNITARY REPRESENTATIONS OF W(INFINITY).}
%Mod. Phys. Lett. {\bf A5}, 2039-2050 (1990)}.

\itemitem{[BK]}{Bakas, I., Kiritsis, E.:
{\it Universal $W$--Algebras in Quantum Field Theory},
Int. J. Mod. {\bf A6}, 2871-2890 (1991)}

\itemitem{[BPRSS]}{Bergshoeff, E., Pope, C., Romans, L., Sezgin, E.,
Shen, X.:
{\it The Super \BR W(infinity) Algebra},
Phys. Lett. {\bf B245}, 447-452 (1990)}

\itemitem{[BPZ]}{Belavin, A., Polyakov, A., Zamolodchikov, A.:
{\it Infinite Conformal Symmetry in Two-Dimensional Quantum Field
Theory},
Nucl. Phys. {\bf B241}, 333-380 (1984)}

\itemitem{[BS]}{Bouwknegt, P., Schoutens, K.:
{\it $W$ Symmetry in Conformal Field Theory},
Phys. Rep. {\bf 223}, 183-276 (1993), and references therein}

\itemitem{[CTZ]}{Cappelli, A., Trugenberger, C., Zemba, G.:
{\it Infinite Symmetry in the Quantum Hall Effect},
Nucl. Phys. {\bf B396}, 465-490 (1993);
{\it Classification of Quantum Hall Universality Classes
by $\Winf$ Symmetry},
Preprint MPI-PH-93-75, November 1993, (hepth/9310181)}

\itemitem{[DLS]}{Douglas, M., Li, K.-K., Staudacher, M.:
{\it Generalized Two--Dimensional QCD},
Preprint LPTENS-94/2, IASSN-HEP-94/3, RU-94-8,
January 1994, (hep-th/ 9401062)}

%\itemitem{[DJKM]}{E. Date, M. Jimbo, M. Kashiwara and T. Miwa,
%{\it Transformation Group for Soliton Equations.}
%Proc. of RIMS Symposium on
%Non-Linear Integrable Systems -
%Classical Theory and Quantum Theory,
%(ed. M. Jimbo and T. Miwa), 39-120, World Sci., 1983}

\itemitem{[DVV]}{Dijkgraaf, R., Verlinde, E., Verlinde, H.:
{\it Loop Equations and Virasoro Constraints in Nonperturbative 2--D
Quantum Gravity},
Nucl. Phys. {\bf B348}, 435-456 (1991)}

\itemitem{[EOTY]}{Eguchi, T., Ooguri, H., Taormina, A., Yang, S.-K.:
{\it Supercomformal Algebras and String Compactification of Manifolds
with $SU(n)$ Holonomy},
Nucl. Phys. {\bf B315}, 193-221 (1989)}

\itemitem{[F]}{Feigin, B.:
{\it The Lie Algebra $gl(\lambda)$ and the
Cohomology of the Lie Algebra of Differential Operators},
Usp. Mat. Nauk {\bf 35}, 157-158 (1988)}

\itemitem{[FFZ]}{Fairlie, D., Fletcher, P., Zachos, C.:
{\it Infinite Dimensional Algebras and a Trigonometric Basis for the
Classical Lie Algebras},
Jour. Math. Phys. {\bf 31}, 1088-1094 (1990)}

\itemitem{[FKN]}{Fukuma, M., Kawai, H., Nakayama, R.:
{\it Continuum Schwinger--Dyson Equations and Universal Structures in
Two--Dimensional Quantum Gravity},
Int. J. Mod. Phys. {\bf A6}, 1385-1406 (1991);
{\it Infinite Dimensional Grassmannian Structure of
Two-Dimensional Quantum Gravity},
Commun. Math. Phys. {\bf 143}, 371-403 (1991)}

\itemitem{[FMS]}{Friedan, D., Martinec, E., Shenker, S.:
{\it Conformal Invariance, Supersymmetry and String Theory},
Nucl. Phys. {\bf B271}, 93-165 (1986)}
% {\it COVARIANT QUANTIZATION OF SUPERSTRINGS.}
% Phys. Lett. {\bf B160}, 55 (1985)}

\itemitem{[G]}{Goeree, J.:
{\it W constraints in 2-D Quantum Gravity},
Nucl. Phys. {\bf B358}, 737-757 (1991)}

\itemitem{[GS]}{Gervais, J., Sakita, B.:
{\it Field Theory Interpretation of Supergauges in Dual Models},
Nucl. Phys. {\bf B34}, 632-639 (1971)}

\itemitem{[GT]}{Gross, D., Taylor, W.:
{\it Twists and Wilson Loops in
the String Theory of Two--Dimensional QCD},
Nucl. Phys. {\bf B400}, 181-208 (1993)}

\itemitem{[IKS]}{Iso, S., Karabali, D., Sakita, B.:
{\it Fermions in the Lowest Landau Level: \BR
Bosonization, $W$ Infinity Algebra, Droplets, Chiral Bosons},
Phys. Lett. {\bf B296}, 143-150 (1992)}

\itemitem{[IM]}{Itoyama, H., Matsuo, Y.:
{\it $\Winf$ type constraints in Matrix Models at Finite $N$},
Phys. Lett. {\bf B262}, 233-239 (1991)}

\itemitem{[IMY]}{Inami,T., Matsuo, Y., Yamanaka, I.:
{\it Extended Conformal Algebras with $N=1$ Supersymmetry},
Phys. Lett. {\bf B215}, 701-705 (1988);
{\it Extended Conformal Algebras with $N=2$ Supersymmetry},
Int. J. Mod. Phys. {\bf A5}, 4441-4468 (1990).}

\itemitem{[KR]}{Kac, V., Radul, A.:
{\it Quasifinite Highest Weight Modules over the Lie Algebra of
Differential Operators on the Circle},
Commun. Math. Phys. {\bf 157}, 429-457 (1993)}
% MIT Mathematics preprint, July 1993, (hepth/9308153)}

\itemitem{[KS]}{Kac, V., Schwarz, A.:
{\it Geometrical Interpretation of the Partition Function of 2--D
Gravity},
Phys. Lett. {\bf B257}, 329-334 (1991)}

\itemitem{[KYY]}{Kawai, T., Yamada, Y., Yang, S.-K.:
{\it Elliptic Genera and $N=2$ Superconformal Field Theory},
Preprint KEK-TH-362, KEK preprint 93-51, June 1993}

\itemitem{[Li]}{Li, W.-L.:
{\it 2--Cocylcles on the Algebra of Differential Operators},
J. Algebra {\bf 122}, 64-80(1989)}

\itemitem{[M]}{Matsuo, Y.:
{\it Free Fields and Quasi-Finite Representation of $\Winf$ Algebra},
Preprint UT-661, December 1993, (hepth/9312192),
to be published in Phys. Lett. B}

\itemitem{[MR]}{Manin, Yu., Radul, A.:
{\it A Supersymmetric Extension of the Kadmtsev--\BR Petviashivili
Hierarchy},
Comm. Math. Phys. {\bf 98}, 65-77 (1985)}

\itemitem{[NS]}{Neveu, A., Shwarz, J.:
{\it Factorizable Dual Model of Pions},
Nucl. Phys. {\bf B31}, 86-112 (1971)}

\itemitem{[O]}{Odake, S.:
{\it Unitary Representations of $W$ Infinity Algebras},
Int. J. Mod. Phys. {\bf A7}, 6339-6355 (1992)}

%\itemitem{[OS]}{S. Odake and T. Sano,
%{W(1) + INFINITY AND SUPERW(INFINITY) ALGEBRAS WITH SU(N) SYMMETRY.}
%Phys. Lett. {\bf B258}, 369-374 (1991)}

\itemitem{[P]}{Park, Q.-H.:
{\it Selfdual Gravity as a Large $N$ Limit of the Two--Dimensional
Nonlinear Sigma Model},
Phys. Lett. {\bf B238}, 287-290 (1991)}

\itemitem{[PRS1]}{Pope, C., Romans, L., Shen, X.:
{\it $W_\infty$ and the Racah-Wiger Algebra},
Nucl. Phys. {\bf B339}, 191-221 (1990)}

\itemitem{[PRS2]}{Pope, C., Romans, L., Shen, X.:
{\it A New Higher Spin Algebra and the Lone Star Product},
Phys. Lett. {\bf B242}, 401-406 (1990)}

\itemitem{[R]}{Ramond, P.:
{\it Dual Theory for Free Fermions},
Phys. Rev. {\bf D3}, 2415-2418 (1971)}

\itemitem{[S]}{Schwarz, A.:
{\it On Solutions to the String Equation},
Mod. Phys. Lett. {\bf A6}, 2713-2726 (1991)}

\itemitem{[SS]}{Schwimmer, A., Seiberg, N.:
{\it Comments on the $N=2$, $N=3$, $N=4$ Superconformal Algebras in
Two Dimensions},
Phys. Lett. {\bf B184}, 191-196 (1987)}

\itemitem{[T]}{Takasaki, K.:
{\it A New Approach to the Self-Dual Yang-Mills Equations},
Comm. Math. Phys. {\bf 94}, 35-59 (1984)}

\itemitem{[UY]}{Ueno, K., Yamada, H.:
{\it Supersymmetric Extension of the Kadomtsev--\BR Petviashvilli
Hierarchy and the Universal Super Grassmann Manifold},
Adv. Stud. Pure. Math. {\bf 16}, 373-426 (1988)}

\itemitem{[YC]}{Yamagishi, K., Chapline, K.:
{\it Induced 4--D Selfdual Quantum Gravity: Affine $W_\infty$
Algebraic Approach},
Class. Quant. Grav. {\bf 8}, 427-446 (1991)}

\itemitem{[Z]}{Zamolodchikov, A.:
{\it Infinite Additional Symmetries in Two--Dimensional Conformal
Quantum Field Theory},
Theor. Math. Phys. {\bf 65}, 347-359 (1985)}

%%%%%%%%%%%%%%%%%%%%%%%%%%%%%%%%%%%%%%%%%%%%%%%%%%%%%%%%%%%%%%%%%%%%%%

\bye